\documentclass[11pt]{article}
\usepackage[utf8]{inputenc}
\usepackage[T1]{fontenc}
\usepackage[OT1]{eulervm}
\usepackage{times}
\usepackage{enumerate}
\usepackage{fixltx2e}
\usepackage{graphicx}
\usepackage{longtable}
\usepackage{float}
\usepackage{setspace}
\usepackage{wrapfig}
\usepackage{soul}
\usepackage{amsmath}
\usepackage{mathtools}
\usepackage{mathrsfs}
\usepackage{textcomp}
\usepackage{marvosym}
\usepackage{wasysym}
\usepackage{latexsym}
\usepackage{amssymb}
\usepackage{hyperref}
\usepackage{graphicx}
\usepackage{subfig}
\usepackage{longtable}
\usepackage{float}
\usepackage{array}
\usepackage[ruled]{algorithm}
\usepackage[noend]{algpseudocode}
\usepackage[table]{xcolor}
\usepackage{arydshln}
\usepackage{cancel}
\usepackage{url}
\usepackage[round, colon]{natbib}
\usepackage[margin=1in]{geometry}
\usepackage[affil-it]{authblk}

\usepackage{color}

\input{./Definitions}

\title{Expandable Factor Analysis}

\author[1]{Sanvesh Srivastava \thanks{\url{sanvesh-srivastava@uiowa.edu}}} 
\author[2]{Barbara E. Engelhardt \thanks{\url{bee@princeton.edu}}}
\author[3]{David B. Dunson \thanks{\url{dunson@duke.edu}}}

\affil[1]{Department of Statistics and Actuarial Science, University of Iowa, 241 Schaeffer Hall, 20 East Washington Street, Iowa City, Iowa 52242, U.S.A.}
\affil[2]{Department of Computer Science, Center for Statistics and Machine Learning, Princeton University, 35 Olden Street, Princeton, New Jersey 08540, U.S.A.}
\affil[3]{Department of Statistical Science, Duke University, Box 90251, Durham, North Carolina 27708, U.S.A.}

\date{\today}

\begin{document}

\maketitle

\begin{abstract}
Bayesian sparse factor models have proven useful for characterizing dependence in multivariate data, but scaling computation to large numbers of samples and dimensions is problematic. We propose expandable factor analysis for scalable inference in factor models when the number of factors is unknown. The method relies on a continuous shrinkage prior for efficient maximum a posteriori estimation of a low-rank and sparse loadings matrix.  The structure of the prior leads to an estimation algorithm that accommodates uncertainty in the number of factors. We propose an information criterion to select the hyperparameters of the prior. Expandable factor analysis has better false discovery rates and true positive rates than its competitors across diverse simulations. We apply the proposed approach to a gene expression study of aging in mice, illustrating superior results relative to four competing methods.
\end{abstract}


\section{Introduction}

Factor analysis is a popular approach to modeling covariance matrices. Letting $k^*$, $p$, and $\Omega$ denote the true number of factors, number of dimensions, and $p \times p$ covariance matrix, factor models set $\Omega = \Lambda \Lambda^T + \Sigma$, where $\Lambda \in \Re^{p \times k^*}$ is the loadings matrix and $\Sigma$ is a diagonal matrix of positive residual variances. To allow computation to scale to large $p$, $\Lambda$ is commonly assumed to be low rank and sparse. These assumptions imply that $k^* \ll p$ and the number of nonzero loadings is small.  A practical problem is that $k^*$ and the locations of zeros in $\Lambda$ are unknown. A number of Bayesian approaches exist to model this uncertainty in $k^*$ and sparsity \citep{Caretal08,KnoGha11}, but conventional approaches that rely on posterior sampling are intractable for large sample size $n$ and dimension $p$.  Continuous shrinkage priors have been proposed that lead to computationally efficient sampling algorithms \citep{BhaDun11}, but the focus is on estimating $\Omega$ with $\Lambda$ treated as a non-identifiable nuisance parameter. Our goal is to build a computationally tractable approach for inference on $\Lambda$ {that models} the uncertainty in $k^*$ and the locations of zeros in $\Lambda$. To do this, we propose a novel shrinkage prior and corresponding class of efficient inference algorithms for factor analysis. 

Penalized likelihood methods provide computationally efficient approaches for point estimation of $\Lambda$ and $\Sigma$. If $k^*$ is known, then many such methods exist \citep{KneSar11, BaiLi12}. Sparse principal components analysis estimates a sparse $\Lambda$ assuming $\Sigma = \sigma^2 I_p$, where $I_p$ is the $p \times p$ identity matrix \citep{Joletal03,ZouHasTib06,SheHua08,WitTibHas09}. The assumptions of spherical residual covariance and known $k^*$ are restrictive in practice.  There are several approaches to {estimating} $k^*$. In econometrics, it is popular to rely on test statistics based on the eigenvalues of the empirical covariance matrix \citep{Ona09,AhnHor13}.  It is also common to fit the model for different choices of $k^*$, and choose the best value based on an information criterion  \citep{BaiNg02}. Recent approaches instead use the trace norm or the sum of column norms of $\Lambda$ as a penalty in the objective function to estimate $k^*$ \citep{CanXu14}.  Alternatively, \citet{RocGeo16} use a spike and slab prior to induce sparsity in $\Lambda$ with an Indian buffet process allowing uncertainty in $k^*$; a parameter-expanded {expectation-maximization} algorithm is then used for estimation.

We propose a Bayesian approach for estimation of a low-rank and sparse $\Lambda$, allowing $k^*$ to be unknown. Our approach relies on a novel multiscale generalized double Pareto prior, inspired by the generalized double Pareto prior for variable selection \citep{ArmDunLee11} and by the multiplicative gamma process prior for loadings matrices \citep{BhaDun11}. The latter approach focuses on estimation of $\Omega$, but does not explicitly estimate $\Lambda$ or $k^*$. The proposed prior leads to an efficient and scalable computational algorithm for obtaining a sparse estimate of $\Lambda$ with appealing practical and theoretical properties. We refer to our method as expandable factor analysis because it allows the number of factors to increase as more dimensions are added and as $p$ increases.

Expandable factor analysis combines the representational strengths of Bayesian approaches with the computational benefits of penalized likelihood methods. The multiscale generalized double Pareto prior is concentrated near low-rank matrices; in particular, a high probability is placed around matrices with rank $O(\log p)$. Local linear approximation of the penalty imposed by the prior equals a sum of weighted $\ell_1$ penalties on the elements of $\Lambda$. This facilitates maximum a posteriori estimation of a sparse $\Lambda$ using {an} extension of the coordinate descent algorithm for weighted $\ell_1$-regularized regression \citep{ZouLi08}. The hyperparameters of our prior are selected using a version of the Bayesian information criterion for factor analysis. Under the theoretical setup for high-dimensional factor analysis in \citet{KneSar11}, we show that the estimates of loadings are consistent, and the estimates of nonzero loadings are asymptotically normal.

\section{Expandable Factor Analysis}

\subsection{Factor analysis model}
\label{sec:factor-analysis-as}

Consider the usual factor model. Let $Y \in \Re^{n \times p}$, $Z \in \Re^{n \times k^*}$, and $E \in \Re^{n \times p}$ be the mean-centered data matrix, latent factor matrix, and residual error matrix, where $Z$ and $E$ are unknown. We use index $i$ for samples, index $d$ for dimensions, and index $j$ for factors. If $\Sigma = \mbox{diag}(\sigma_{1}^2,\ldots,\sigma_{p}^2)$ is the residual error variance matrix, then the factor model for $y_{id}$ is
\begin{align}
  y_{id} = \sum_{j=1}^{k^*} z_{ij} \lambda_{dj} + e_{id}, \quad z_{ij} \sim N(0, 1), \quad e_{id} \mid \sigma_{d}^2 \sim N(0, \sigma_{d}^2), \label{eq:scal}
\end{align}
where $z_{ij}$ and $e_{id}$ are independent $(i=1, \ldots, n; \; j=1, \ldots, k^*; \; d=1, \ldots, p)$. Equivalently, 
\begin{align}
  y_i = \Lambda z_i + e_i, \quad y_i = (y_{i1}, \ldots, y_{ip})^T, \quad z_i = (z_{i1}, \ldots, z_{ik^*})^T, \quad e_{i} = (e_{i1}, \ldots, e_{ip})^T \label{mdl-1} 
\end{align}
for sample $i$ and $\text{cov}(y_i) = \Lambda \Lambda^T + \Sigma$ $(i = 1, \ldots, n)$. Similarly, model \eqref{eq:scal} reduces to regression in the space of latent factors
\begin{align}
   y_d = Z \lambda_d + e_d, \quad  y_d = (y_{1d}, \ldots, y_{nd})^T, \quad \lambda_d = (\lambda_{d1}, \ldots, \lambda_{dk^*})^T, \quad e_{d} = (e_{1d}, \ldots, e_{nd})^T  \label{fa-reg}
\end{align}
for dimension $d$ $(d=1, \ldots, p)$. Unlike usual regression, the design matrix $Z$ in \eqref{fa-reg} is unknown. 

Penalized estimation of $\Lambda$ is typically based on \eqref{mdl-1} or \eqref{fa-reg}. {The loss is estimated as the regression-type squared error after imputing $Z$ using the eigen-decomposition of the empirical covariance matrix $Y^TY/n$ or an {expectation-maximization} algorithm. The choice of penalty on $\Lambda$ presents a variety of options.} If the goal is to select factors that affect any of the $p$ variables, then the sum of column norms of $\Lambda$ can be used as a penalty; a recent example is the group bridge penalty, $\sum_{j=1}^{k} (\sum_{d=1}^p \lambda_{dj}^2 / p)^{\alpha}$, where $0 < \alpha < 1/2$ and $k$ is an upper bound on $k^*$. The selected factors correspond to the non-zero columns of the estimated $\Lambda$ \citep{CanXu14}. To further obtain element-wise sparsity, a non-concave variable selection penalty can be applied to the elements in $\Lambda$. The estimate of $\Lambda$  depends on the choice of criterion for selecting the tuning parameters \citep{HirYam13}. 

Our expandable factor analysis differs from this typical approach in several important ways.  We start from a Bayesian perspective, and place a prior on $\Lambda$ that is structured to allow uncertainty in $k^*$ while shrinking towards loadings matrices with many zeros and $k^* \ll p$. If $k$ is an upper bound on $k^*$, then the prior is designed to automatically allow a slow rate of growth in $k$ as the number of dimensions $p$ increases by concentrating in neighborhoods of matrices with rank bounded above by $k=O(\log p)$.  To our knowledge, this is a unique feature of our approach, justifying {its name}.  Expandability is an appealing characteristic, as more factors should be needed to accurately model the dependence structure as the dimension of the data increases.

\subsection{Multiscale generalized double Pareto prior}
\label{sec:multi-scale-gener}

We would like to design a prior on $\Lambda$ such that maximum a posteriori estimates of $\Lambda$ have the following four characteristics:
\begin{itemize}
\item[(a)] the estimate of a loading with large magnitude should be nearly unbiased;
\item[(b)] a thresholding rule, such as soft-thresholding, is used to estimate the loadings so that loadings estimates with small magnitudes are automatically set to zero;
\item[(c)] the estimator of any loading is continuous in the data to limit instability; and
\item[(d)] the $\ell_2$ norm of the $i$th column of the estimated $\Lambda$ does not increase as $i$ increases. 
\end{itemize}
The first three properties are related to non-concave variable selection \citep{FanLi01}. Properties (b) and (d) together ensure existence of a {column index} after which all estimated loadings are identically zero. Automatic relevance determination and multiplicative gamma process priors satisfy (d) but fail to satisfy (b). {No existing prior for loadings matrix satisfies} properties (a)--(d) simultaneously \citep{Caretal08,BhaDun11,KnoGha11}.

In order to satisfy these four properties and obtain a computationally efficient inference procedure, it is convenient to start with a prior for a loadings matrix $\Lambda \in \Re^{p \times \infty}$ having infinitely many columns; in practice, all of the elements will be estimated to be zero after a finite column index {that corresponds} to the estimated number of factors.  \citet{BhaDun11} show that the set of loadings matrices $\Lambda \in \Re^{p \times \infty}$ that leads to well-defined covariance matrices is
\begin{align*}
  \Thetabb = \bigg\{ \Lambda: \underset{1 \leq d \leq p}{\max} \sum_{j=1}^{\infty} \lambda_{dj}^2 < \infty\bigg\}.
\end{align*}
We propose a multiscale generalized double Pareto prior for $\Lambda$ having support on $\Thetabb$.  This prior is constructed to concentrate near low-rank matrices, placing high probability around matrices with rank at most $k=O(\log p)$. 

The multiscale generalized double Pareto prior on $\Lambda$ specifies independent generalized double Pareto priors on $\lambda_{dj}$ ($d=1, \ldots, p$; $j=1, \ldots, \infty$) so that the density of $\Lambda$ is
\begin{align}
  p_{\mathrm{mgdP}}(\Lambda) = \prod_{d=1}^p \prod_{j=1}^{\infty} p_{\mathrm{gdP}}(\lambda_{dj} \mid \alpha_j, \eta_j), \quad p_{\mathrm{gdP}}(\lambda_{dj} \mid \alpha_j, \eta_j) = \frac{\alpha_j}{2 \eta_j} \bigg(1 + \frac{|\lambda_{dj}|}{\eta_j} \bigg)^{-(\alpha_j + 1)}, \label{mdl-3}
\end{align}
where $p_{\mathrm{gdP}}(\cdot \mid \alpha_j, \eta_j)$ is the generalized double Pareto density with parameters $\alpha_j$ and $ \eta_j$ \citep{ArmDunLee11}. This prior on $\lambda_{dj}$ ensures that properties (a)--(c) are satisfied. Property (d) is satisfied by choosing parameter sequences $\alpha_j$ and $\eta_j$ ($j=1, \ldots, \infty$) such that two conditions hold: the prior measure $P_{l}$ on $\Thetabb$ has density $p_{\mathrm{mgdP}}$ in \eqref{mdl-3} and $P_{l}$ has $\Thetabb$ as its support. These conditions hold for the form of $\alpha_j$ and $\eta_j$ ($j=1, \ldots, \infty$) specified by the following lemma. 
\begin{lemma}\label{lem:gdp-1}
  If $\alpha_j > 2$, $\eta_j/\alpha_j = O(j^{-m})$ ($j=1, \ldots, \infty$) and $m > 1/2$, then $P_{l} (\Thetabb)= 1$.
\end{lemma}
The proof is in the Supplementary Material, with the other proofs. 

As in \citet{BhaDun11}, we truncate to a finite number of columns for tractable computation.  This truncation is accomplished by mapping $\Lambda \in \Thetabb$ to $\Lambda^{k} \in \Thetabb$, with $\Lambda^{k}$ retaining the first $k$ columns of $\Lambda$.  
The choice of $k$ is such that $\Omega^{k} = \Lambda^{k} \Lambda^{{k}^T} + \Sigma$ is arbitrarily close to $\Omega = \Lambda \Lambda^T + \Sigma$, where distance between $\Omega^{k}$ and $\Omega$ is measured using the $\ell_{\infty}$ norm of their element-wise difference.  In addition, for computational convenience, we assume that the hyperparameters $\alpha_j$ and $\eta_j$ ($j=1, \ldots, \infty$) are analytic functions of parameters $\delta$ and $\rho$, respectively, with these functions satisfying the conditions of Lemma \ref{lem:gdp-1}.  

The following lemma defines the form of $\alpha_{j}$ and $\eta_{j}$  ($j=1, \ldots, \infty$) in terms of $\delta$ and $\rho$.
\begin{lemma}\label{lem:gdp-2}
  If $\delta > 2$, $\rho > 0$, $\alpha_j(\delta) = \delta^j$, and $\eta_j(\rho) = \rho$ ($j = 1, \ldots, \infty$), then $P_{l} (\Thetabb) = 1$, where $P_{l}$ has density $p_{\mathrm{mgdP}}$ in \eqref{mdl-3} with hyperparameters $\alpha_j(\delta)$ and $\eta_j(\rho)$ ($j = 1, \ldots, \infty$). Furthermore, given $\epsilon > 0$, there exists a positive integer $k(p, \delta, \epsilon)= O\left( \log^{-1} \delta \log \frac{p}{\epsilon^2}\right)$ for every $\Omega$ such that for all $r \geq k$, $\alpha_j(\delta)=\delta^j$, $\eta_j(\rho)=\rho$, ($j=1, \ldots, r$), and $\Omega^{r} = \Lambda^{r} \Lambda^{r^T} + \Sigma$, pr$\{\Omega^{r} \, | \, d_{\infty}(\Omega, \Omega^{r}) < \epsilon\} > 1 - \epsilon$, where $d_{\infty}(A, B) = \underset{1 \leq i,j \leq p}{\max}|a_{ij} - b_{ij}|$. 
\end{lemma}
The penalty imposed on the loadings by the prior has exponential growth in terms of $\delta$ as the column index increases. This property of the prior ensures that all the loadings are estimated to be zero after a finite column index, which corresponds to the estimated number of factors.

\section{Estimation algorithm}
\label{sec:estim-fact-load}

\subsection{Expectation-maximization algorithm}
\label{sec:estim-param-expans}

We rely on an adaptation of the expectation-maximization algorithm {to estimate $\Lambda$ and $\Sigma$}.  
Choose a positive integer $k$ of order $\log p$ as the upper bound on $k^*$; the estimate of the number of factors will be less than or equal to $k$. Results are not sensitive to the choice of $k$ due to the properties of the multiscale generalized double Pareto prior, provided $k$ is sufficiently large. If $k$ is too small, then the estimated number of factors will be equal to the upper bound, suggesting to increase this bound. Given $k$, define $\alpha_j(\delta)$ and $\eta_j(\rho)$ for $j=1, \ldots, k$ as in Lemma \ref{lem:gdp-2}, with $\delta>2$ and $\rho>0$ being pre-specified constants.

We present the objective function as a starting point for developing the coordinate descent algorithm and provide derivations in the Supplementary Material. Let $F^{(t)} = n^{-1} E( Z^T Z \mid Y, \Lambda^{(t)}, \Sigma^{(t)})$ and $L^{(t)} = n^{-1} E( \sum_{i=1}^n y_i z_i^T \mid Y, \Lambda^{(t)}, \Sigma^{(t)})$, where the superscript $(t)$ denotes an estimate at iteration $t$ and  $E(\cdot \mid Y, \Lambda^{(t)}, \Sigma^{(t)})$ denotes the conditional expectation given $Y$, $\Lambda^{(t)}$, and $\Sigma^{(t)}$ based on 
\eqref{eq:scal}. The objective function for parameter updates in iteration $(t+1)$ is
\begin{align}
  \underset{\begin{subarray}{c}\lambda_{d}, \, \sigma_{d}^2\\
      d=1, \ldots, p
    \end{subarray}} {\argmin}\;  \sum_{d=1}^p \bigg( \frac{n+2}{2npk} \log \sigma_{d}^2 &+ \frac{\| w_{d}^{(t)} - X^{(t)} \lambda_d\|^2 - w_{d}^{(t)^T} w_{d}^{(t)} + (Y^TY/n)_{dd}}{2 pk\sigma^2_{d}} 
   \nonumber \\
                                                                                      &+ \bigg[ \sum_{j=1}^k \frac{\alpha_j(\delta) +1}{npk} \log \bigg\{ 1 + \frac{|\lambda_{dj}|}{\eta_j(\rho)} \bigg\} \bigg] \bigg),   \label{obj}
\end{align}
where $ X^{(t)} = F^{(t){1/2}}$ and $w_{d}^{(t)} = F^{(t){-1/2}} l_d^{(t)}$ ($d = 1, \ldots, p$).

\subsection{Estimating parameters using a convex objective function}

The objective (\ref{obj}) is written as a sum of $p$ terms. The $d$th term corresponds to the objective function for the regularized estimation of the $d$th row of the loadings matrix, $\lambda_d^T$, with a specific form of log penalty on $\lambda_d$ \citep{ZouLi08}. Local linear approximation at $\lambda_{dj}^{(t)}$ of the log penalty on $\lambda_{dj}$ in \eqref{obj} implies that each row of $\Lambda$ is estimated separately at iteration $(t+1)$:
\begin{align}
  \lambda_d^{\mathrm{lla}{(t+1)}} = \underset{\lambda_d} {\argmin} \, \frac{\| w_{d}^{(t)} - X^{(t)} \lambda_d\|^2}{2pk\sigma^{2{(t)}}_{d}} + \sum_{j=1}^k \frac{ \alpha_j(\delta) + 1}{npk \big\{\eta_j(\rho) + |\lambda_{dj}^{(t)}| \big\}} |\lambda_{dj}|, \quad (d=1, \ldots, p). \label{lla-obj}
\end{align}
This problem corresponds to regularized estimation of regression coefficients $\lambda_d$ with $w_d^{(t)}$ as the response, $X^{(t)}$ as the design matrix, $\sigma^{2{(t)}}_{d}$ as the error variance, and a weighted $\ell_1$ penalty on $\lambda_d$. 

The solution of \eqref{lla-obj} is found using block coordinate descent. Let column $j$ of $F$ and row $d$ of $\Lambda$ without the $j$th element be written $f_{(-j),j}$ and $\lambda_{d, (-j)}^T$. Then the update to estimate $\lambda_d^{\mathrm{lla}}$ is
\begin{align}
   \lambda^{\mathrm{lla}{(t+1)}}_{dj} = \frac{\sgn(\widetilde \lambda_{dj}^{(t)})} {f_{jj}^{(t)}} \left(|\widetilde \lambda_{dj}^{(t)}| - c_{dj}^{(t)} \right)_{+}, \quad c_{dj}^{(t)} = \frac{\sigma^{2{(t)}}_{d}  \big\{ \alpha_j(\delta) + 1\big\} } {n  \big\{ \eta_j(\rho) + |\lambda_{dj}^{(t)}| \big\} }, \quad (j=1, \ldots, k),\label{lla-sol}
\end{align}
where $\widetilde \lambda_{dj}^{(t)} = l_{dj}^{(t)} - \lambda_{d, (-j)}^{\mathrm{lla}{(t)^T}} f^{(t)}_{(-j), j}$  and $(x)_{+}=\max(x, 0)$. Fix $\lambda_d$ at $\lambda_d^{\mathrm{lla}{(t+1)}}$ in (\ref{obj}) to update $\sigma_{d}^2$  in iteration $(t+1)$ as
\begin{align}
  \sigma_{d}^{2{{(t+1)}}}  &= \frac{n}{n+2} \big\{ (Y^TY/n)_{dd} +  \lambda^{\mathrm{lla}{(t+1)^T}} F^{(t)} \lambda^{\mathrm{lla}{(t+1)}}
-2 l_d^{(t)^T} \lambda^{\mathrm{lla}{(t+1)}} \big\}. \label{lla-sig}
\end{align}
If any root-$n$ consistent estimate of $\lambda_{dj}$ is used instead of $\big|\lambda_{dj}^{(t)}\big|$ in (\ref{lla-obj}), then it acts as a warm starting point for the estimation algorithm. This leads to a consistent estimate of $\lambda_{dj}$ in one step of coordinate descent \citep{ZouLi08}. An implementation of this approach for known values of $\delta$ and $\rho$ is summarized in steps (i)--(iv) of Algorithm \ref{xfa-alg} using R \citep{R16} package glmnet \citep{Frietal10}.

The estimate of $\Lambda$ obtained using (\ref{lla-sol}) satisfies properties (a)--(d) described earlier. The adaptive threshold $c_{dj}^{(t)}$ in (\ref{lla-sol}) ensures that property (a) is satisfied. The soft-thresholding rule to estimate $\lambda_{dj}$ ensures that property (b) is satisfied. The local linear approximation (\ref{lla-obj}) has continuous first derivatives in the parameter space excluding zero, so property (c) is also satisfied \citep{ZouLi08}. The $\Lambda$ estimate satisfies property (d) due to the structured penalty imposed by the prior. 

{
We comment briefly on the choice of prior and uncertainty quantification. We build on the generalized double Pareto prior instead of other shrinkage priors not only because the estimate of $\Lambda$ satisfies properties (a)--(d), but also because local linear approximation of the resulting penalty has a weighted $\ell_1$ form. We exploit this for efficient computations and use a warm starting point to estimate a sparse $\Lambda$ in one step using Algorithm \ref{xfa-alg}. Uncertainty estimates of the nonzero loadings are obtained from Laplace approximation, and the remaining loadings are estimated as 0 without uncertainty quantification. 
}

{\small
\begin{enumerate}[]
  \begin{algorithm}[!h]
    \caption{Estimation algorithm for expandable factor analysis} \label{xfa-alg}
  \item {\bf Notation}: 
    \begin{enumerate}[1.]
    \item $\diag(A)$ is the diagonal matrix containing diagonal elements of a symmetric matrix $A$. 
    \item Chol($A$) is the upper triangular Cholesky factorization of  a symmetric positive definite matrix $A$. 
    \item $\mathrm{bdiag}(A_1, \ldots, A_p)$ is a block diagonal matrix with $A_1, \ldots, A_p$ forming the diagonal blocks.
    \item $\text{vec}(A) = (a_1^T, \ldots, a_d^T)^T \in \Re^{cd \times 1}$, where $A \in \Re^{c \times d}$.
    \end{enumerate}
  \item {\bf Input}: 
    \begin{enumerate}[1.]
    \item Data $Y \in \Re^{n \times p}$ and upper bound $k=O(\log p)$ on the rank of the loadings matrix.
    \item The $\delta$-$\rho$ grid with $RS$ grid indices ($\delta_1 < \cdots < \delta_R$; $\rho_1 < \cdots < \rho_S$). 
    \end{enumerate}
  \item {\bf Do}: 
    \begin{enumerate}[1.]
    \item Center data about their mean $\hat y_{ij} = y_{ij} - \frac{1}{n}\sum_{m=1}^n y_{mj}$ ($i = 1, \ldots, n$; $ j = 1, \ldots, p$).
    \item Let $S_{\hat y \hat y} = \hat Y^T \hat Y/n$, then estimate eigenvalues and eigenvectors of $S_{\hat y \hat y}$: $\lambdah_1, \ldots, \lambdah_p$ and $\widehat \psi_1, \ldots, \widehat \psi_p$.
    \item Define $\Lambda^0 $ to be the matrix $\{ (\lambdah_1)^{1/2} \psih_{1}, \cdots, (\lambdah_k)^{1/2} \psih_{k} \}$.
    \item Begin estimation of $\Lambda$, $\Sigma$, and $\pi$ across the $\delta$-$\rho$ grid:
      \begin{enumerate}[]        
      \item For $r=1, \ldots, R$
      \item \quad For $s=S, \ldots, 1$        
      \item \quad \quad (i) Define $\alpha_j = \delta_r^j$, $\eta_j = \rho_s p^{1/2}$ if $n \leq p$, and $\eta_j = \rho_s$ if $n> p$ ($j=1, \ldots, k$).
      \item \quad \quad (ii) Initialize the following statistics required in \eqref{lla-sol}:
      \begin{align*}
        &\Sigma^0 = \diag(S_{\hat y \hat y} - \Lambda^{0} \Lambda^{0^T}),\quad \Omega^0 = \Lambda^{0} \Lambda^{0^T} + \Sigma^{0}, \quad G^{0} = \Omega^{0^{-1}} \Lambda^{0}, \quad L^{0}= S_{\hat y \hat y} G^{0},\\
        &\Delta^{0} = I_k - \Lambda^{0^T}G^{0},\quad F^{0} = \Delta^{0} + G^{0^T} S_{\hat y \hat y} G^{0}, \quad R^0 = \text{Chol}(F^0).
      \end{align*}
      \item \quad \quad (iii) Define $X \in \Re^{pk \times pk}$, $w \in \Re^{pk \times 1}$, $y \in \Re^{pk \times 1}$, and $v \in \Re^{pk \times 1}$ required to solve \eqref{lla-obj}:
      \begin{align*}
        X &= \mathrm{bdiag}(\underbrace{R^0, \ldots, R^0}_{p \text{ times}}),\quad
        w = \{ \underbrace{{(\Sigma^0)_{11}^{{-1}}}, \ldots, {(\Sigma^0)_{11}^{{-1}}}}_{k \text{ times}}, \ldots, \underbrace{{(\Sigma^0)_{pp}^{{-1}}}, \ldots, (\Sigma^0)_{pp}^{{-1}} }_{k \text{ times}} \}, \\
        y &= \text{vec} \left( R^{0^{-1}} L^{0^T} \right), \quad v = \tfrac{1}{npk}\left(\tfrac{\alpha_1 + 1}{\eta_1 + |\lambda^0_{11}|}, \ldots, \tfrac{\alpha_k + 1}{\eta_k + |\lambda^0_{1k}|}, \ldots, \tfrac{\alpha_1 + 1}{\eta_1 + |\lambda^0_{p1}|}, \ldots, \tfrac{\alpha_k + 1}{\eta_k + |\lambda^0_{pk}|} \right).
      \end{align*}
      \item \quad \quad (iv) Estimate $\Lambda^{\text{lla}}$ in \eqref{lla-sol} and $\Sigma^{\text{lla}}$ in \eqref{lla-sig}  using {R} package {glmnet}  in three steps:
      \item \quad \quad \quad - {{result} $\leftarrow$ {glmnet}}({x} = $X$, {y} = $y$, {weights} = $w$, {intercept} = {FALSE}, {standardize} = {FALSE}, 
      \item \qquad \qquad \qquad \qquad \qquad \quad {penalty.factor} = $v / \sum_{j=1}^{pk} v_j $).
      \item \quad \quad \quad - $\text{vec} \left(\Lambda^{\text{lla}^T} \right) \leftarrow $ {coef}({result}, {s = }$\sum_{j=1}^{pk} v_j$, {exact = TRUE}) {[-1, ]}.
      \item \quad \quad \quad - $(\Sigma^{{\text{lla}}})_{dd} = n / (n+2) \big\{ (S_{\hat y \hat y})_{dd} +  \lambda_d^{{\text{lla}}^T} F^{0} \lambda_d^{\text{lla}} -2 l_d^{0^T} \lambda_d^{\text{lla}} \big\}$  ($d = 1, \ldots, p$).
      \item  \quad \quad (v) Set $\Lambda^{(r,s)} = \Lambda^{\text{lla}}$, $\Sigma^{(r,s)} = \Sigma^{\text{lla}}$, $\Lambda^0 = \Lambda^{\text{lla}}$, and estimate posterior weight $\pi^{(r, s)}$ in \eqref{bma-3}.
      \item \quad End for.
      \item \quad Set $\Lambda^0 = \Lambda^{(r, S)}$.
      \item  End for.
      \end{enumerate}      
    \item Obtain grid index $(\hat r, \hat s)$ for the estimate of $(\delta, \rho)$, where $\pi^{(\hat r, \hat s)} = \underset{(r, s)}{\max} \, \pi^{(r, s)}.$ 
    \end{enumerate}
    \item {\bf Return}: 
      \begin{enumerate}[1.]
      \item $\Lambda^{( \hat r, \hat s)}$, $\Sigma^{(\hat r,\hat s)}$, and $\Mcal^{(\hat r, \hat s)} = \{(d, j) : \lambda^{(\hat r, \hat s)}_{dj} \neq 0, \, d=1, \ldots, p, \, j = 1, \ldots, k \}$.
      \end{enumerate}
  \end{algorithm}
\end{enumerate}
}%

\subsection{Root-$n$ consistent estimate of $\lambda_{dj}$}
\label{sec:root-lamdb}

The root-$n$ consistent estimate of $\lambda_{dj}$ exists under Assumptions A.0--A.4 given in the appendix. If $\lambdah_d$ and $\psih_{d}$ ($d=1, \ldots, p$) are the eigenvalues and eigenvectors of the empirical covariance matrix $Y^TY/n$, then $\sum_{d=1}^p \lambdah_d \psih_{d} \psih_{d}^T$ is the eigen decomposition of  $Y^TY/n$. It is known that $({\lambdah_{d}})^{1/2} \psih_{dj}$ is a root-$n$ consistent estimator of $\lambda_{dj}$ if $p$ is fixed and $n \goesto \infty$. If $n \rightarrow \infty$, $n \leq p \rightarrow \infty$, and $\log p /n \rightarrow 0$, then $p^{-1/2} {({\lambdah_d})^{1/2}} \psih_{dj}$ is a root-$n$ consistent estimator of $p^{-1/2}{\lambda_{dj}}$; see the Supplementary Material for a proof. Scaling by $p^{1/2}$ is required because the largest eigenvalue of $\Omega$ tends to infinity as $p \goesto \infty$ \citep{KneSar11}. This scaling does not change our estimation algorithm for $\lambda_{dj}$ in \eqref{lla-sol}, except $\eta_j(\rho)$ is changed to {$\eta_j(\rho) p^{1/2} $} ($j=1, \ldots, k$). 

\subsection{Bayesian information criterion to select $\delta$ and $\rho$}
\label{sec:esim-lamdb-using}

The parameter estimates in \eqref{lla-sol} and \eqref{lla-sig} depend on the hyperparameters through $\delta$ and $\rho$, both of which are unknown. To estimate $\delta$ and $\rho$, we use a grid search. Let $\delta_1 < \cdots < \delta_R$ and $\rho_1 < \cdots < \rho_S$ form a $\delta$-$\rho$ grid. If $(\delta_r, \rho_s)$ is the value of ($\delta$, $\rho$) at grid index $(r, s)$, then  $\alpha_{j}(\delta_r)$ and $\eta_{j}(\rho_s)$ $(j=1, \ldots, k)$ are the hyperparameters of our prior defined using Lemma \ref{lem:gdp-2}, and $\Lambda^{(r, s)}$ and $\Sigma^{(r, s)}$ are the parameter estimates based on this prior. Algorithm \ref{xfa-alg} first estimates $\Lambda^{(r, s)}$ and $\Sigma^{(r, s)}$ for every $(r, s)$ by choosing warm starting points and then estimates ($\delta$, $\rho$) using all the estimated $\Lambda$ and $\Sigma$. These two steps in the estimation of  ($\delta$, $\rho$)  are described next.

The structured penalty imposed by our prior implies that $\Lambda^{(1, S)}$ has the maximum number of nonzero loadings. Algorithm \ref{xfa-alg} exploits this structure by first estimating $\Lambda^{(1, S)}$ and then other loadings matrices along the $\delta$-$\rho$ grid by successively thresholding nonzero loadings in $\Lambda^{(1, S)}$ to 0. Let $\Mcal^{(r, s)} = \{(d, j) : \lambda_{dj}^{(r, s)} \neq 0, \, d=1, \ldots, p, \, j = 1, \ldots, k \}$ be the set that contains the locations of nonzero loadings in $\Lambda^{(r, s)}$. The estimation path of Algorithm \ref{xfa-alg} across the $\delta$-$\rho$ grid is such that $\Mcal^{(r, 1)} \subseteq \cdots \subseteq \Mcal^{(r, S)}$ ($r=1, \ldots, R$) and $\Mcal^{(R, S)} \subseteq \cdots \subseteq \Mcal^{(1, S)}$.

After the estimation of $\Lambda^{(r, s)}$ and $\Sigma^{(r, s)}$ $(r=1, \ldots, R;\; s = 1, \ldots, S)$, $(\delta, \rho)$ is set to $(\delta_{\hat r}, \rho_{\hat s})$ if $\Mcal^{(\hat r, \hat s)}$ has the maximum posterior probability. Let $|A|$ be the cardinality of set $A$. Given $(\delta_{r}, \rho_{s})$, there are $pk \choose \vert \Mcal^{(r,s)} \vert$ loadings matrices that have $\vert \Mcal^{(r,s)} \vert$ nonzero loadings but differ in the locations of nonzero loadings. Assuming that each of these matrices is equally likely to represent the locations of nonzero loadings in the true loadings matrix, the prior for $\Mcal^{(r,s)}$ is
\begin{align}
  \label{bma-2}
  \text{pr}( \Mcal^{(r,s)} \mid \delta_r,\rho_s) \propto {pk \choose \vert \Mcal^{(r,s)} \vert}^{-1},  \quad (r=1, \ldots, R; \; s = 1, \ldots, S). 
\end{align}
Let $\pi^{(r, s)}$ be the posterior probability of $\Mcal^{(r, s)}$. Then an asymptotic approximation to $-2 \log \pi^{(r,s)}$ is
\begin{align}
  \label{bma-3}
    -2 \log f(Y, \Lambda^{(r, s)} \mid \delta_r, \rho_s) + \vert \Mcal^{(r,s)} \vert \, \log n + 2 \, \vert \Mcal^{(r,s)} \vert \, \log (p k)
\end{align}
if terms of order smaller than $\log n + \log p$ are ignored, where $f(Y, \Lambda \mid \delta_r, \rho_s)$ is the joint density of $Y$ and $\Lambda$ based on \eqref{eq:scal}. The first term on the right in \eqref{bma-3} measures the goodness-of-fit, and the last two terms penalize complexity of a factor model with $n$ samples and $pk$ loadings with the locations of nonzero loadings in $\Mcal^{(r,s)}$. Theorem \ref{thm-3} in the next section shows that $- 2 \log \pi^{(r, s)}$ and $\textsc{ebic}_{\gamma}(\Mcal^{(r, s)})$ have the same asymptotic order under certain regularity assumptions, where $\textsc{ebic}_{\gamma}$ is the extended Bayesian information criteria of \citet{CheChe08} and
$0 \leq \gamma \leq 1$ is an unknown constant. The analytic forms of $- 2 \log \pi^{(r, s)}$ and $\textsc{ebic}_{\gamma}(\Mcal^{(r, s)})$ are the same when $\gamma = \text{0$\cdot$5}$ and terms of order smaller than $\log n + \log p$ are ignored, so we use $\textsc{ebic}_{\text{0$\cdot$5}}$ for estimating $\Mcal^{(\hat r, \hat s)}$ in our numerical experiments.

\section{Theoretical properties}
\label{sec:theor-prop}

Let $\Lambda_n^{\mathrm{lla}}$ and $\Sigma_n^{\mathrm{lla}}$ be the fixed points of $\Lambda^{\mathrm{lla}{(t)}}$ and $\Sigma^{\mathrm{lla}{(t)}}.$  The updates (\ref{lla-sol}) and (\ref{lla-sig}) define the map $g:\theta^{(t)} \mapsto  \theta^{(t+1)}$, where $\theta  = (\Lambda, \Sigma)$. The following theorem shows that our estimation algorithm retains the convergence properties of the expectation-maximization algorithm.
\begin{theorem} \label{thm-1}
  If $\Lcal(\theta)$ represents the objective \eqref{obj}, then $\Lcal(\theta)$ does not decrease at every iteration.  Let $Q$ be the local linear approximation of \eqref{obj}. Assume that $Q(\theta) = Q\{g(\theta)\}$ only for stationary points of $Q$, then the sequence $\{ \theta^{(t)} \}_{t=1}^{\infty}$ converges to its stationary point $\theta^{\mathrm{lla}}_n$.
\end{theorem}

Let $\Lambda^*$ and $\Sigma^*$ be the true loadings matrix and residual variance matrix. {We define $\lambda_{dj}^* = 0$ ($d=1, \ldots, p$; $j = k^* + 1, \ldots, k$) and express $\Lambda^*$ as having $k$ columns}. The locations of true nonzero loadings are in the set $\Mcal^* = \{(d, j) : \lambda_{dj}^{*} \neq 0, \, d=1, \ldots, p, \, j = 1, \ldots, k \}$. Let $\hat \Lambda$ and $\hat \Sigma$ be the estimate of $\Lambda$ and $\Sigma$ obtained using our estimation algorithm for a specific choice of $\alpha_j(\delta)$ and $\eta_j(\rho)$ ($j=1, \ldots, k$), then $\hat \Mcal = \{(d, j) : \hat \lambda_{dj} \neq 0, \, d=1, \ldots, p, \, j = 1, \ldots, k \}$ is an estimator of $\Mcal^*$. If $\hat \lambda = \text{vec}(\hat \Lambda^T)$ and  $\lambda^* = \text{vec}(\Lambda^{*^T})$, then 
$\hat \lambda_{A}$ and $\lambda^*_{A}$ retain elements of $\hat \lambda$ and  $\lambda^*$ with indices in the set $A$. The following theorem specifies the asymptotic properties of $\hat \Lambda$, $\hat \Sigma$, and $\hat \Mcal$. 
\begin{theorem} \label{thm-2}
  If Assumptions A.0--A.6 given in the appendix hold and $n \rightarrow \infty$, $n \leq p \rightarrow \infty$, and $\log p / n \rightarrow 0$, then for any $d=1, \ldots, p$ and $j=1, \ldots, k$,
  \begin{enumerate}
  \item $\hat \lambda_{dj}$, $\hat \sigma_d^2$, and $\hat \Mcal$ are consistent estimators of $\lambda_{dj}^*$, $\sigma_d^{2^*}$, and $\Mcal^{*}$, respectively; and
  \item $n^{1/2} (\hat \lambda_{\Mcal^*}  - \lambda^*_{\Mcal^*}) \rightarrow N_{|\Mcal^*|}(0, C_{*})$ and $n^{1/2} (\hat \sigma_d^2 - \sigma^{2*}_d) \rightarrow N(0, c_{*})$ in distribution, where $C_*$ is a $\left|\Mcal^*\right| \times \left|\Mcal^*\right|$ symmetric positive definite matrix and $c_* > 0$.
  \end{enumerate}
\end{theorem}

Theorem \ref{thm-2} holds for any multiscale generalized double Pareto prior with hyperparameters $\alpha_j(\delta)$ and $\eta_j(\rho)$ ($j=1, \ldots, k$) that satisfies Assumption A.5. In practice, the estimate of $\Lambda$ depends on the choice of $\delta$ and $\rho$. Restricting the search to the hyperparameters indexed along the $\delta$-$\rho$ grid, Algorithm \ref{xfa-alg} sets the values of the hyperparameters to $\alpha_j(\delta_{\hat r})$ and $\eta_j(\rho_{\hat s})$ ($j=1, \ldots, k$), where $\pi^{( r,  s)}$ achieves its maximum at grid index $(\hat r, \hat s)$. 
The following theorem justifies this method of selecting hyperparameters and shows the asymptotic relationship between $- 2 \log \pi^{(r, s)}$ and $\textsc{ebic}_{\gamma}(\Mcal^{(r, s)})$. 
 \begin{theorem}\label{thm-3} 
   Suppose the generalized double Pareto prior with hyperparameters defined using $(\delta_*, \rho_*)$ leads to estimation of $\Mcal^*$. Let $\Mcal \neq \Mcal^*$ be another set that contains the locations of nonzero loadings in an estimated $\Lambda$ for a given $(\delta, \rho)$. Define $\pi_{\Mcal} = \text{pr}(\Mcal \mid Y)$ and $\pi_{\Mcal^*} = \text{pr}(\Mcal^* \mid Y)$. If Assumptions A.0--A.7  given in the appendix hold, then for any $\Mcal$ such that $|\Mcal| \in \{1, \ldots, pk\}$,
  \begin{enumerate}
  \item $- 2 \log \pi_{\Mcal} / \textsc{ebic}_{\gamma}(\Mcal) \rightarrow 1$ in probability as $n \rightarrow \infty$; and
  \item pr$\{\max (\pi_{\Mcal} : \Mcal \neq \Mcal^*) < \pi_{\Mcal^*} \} \rightarrow 1$ as $n \rightarrow \infty$. 
  \end{enumerate}
\end{theorem}
Let $(\delta_{r^*}, \rho_{s^*})$ be a point on the $\delta$-$\rho$ grid that leads to estimation of $\Mcal^*$. Then Theorem \ref{thm-3} shows that Algorithm \ref{xfa-alg} selects $\Mcal^*$ with probability tending to 1 because $\pi^{( r^*,  s^*)}$ will be larger than any $\pi^{( r,  s)}$, where $( r,  s)$ is such that $\Mcal^{( r,  s)} \neq \Mcal^*$. 

\section{Data Analysis}
\label{sec:experiments}

\subsection{Setup and comparison metrics}
We compared our method with those of \citet{CanXu14}, \citet{HirYam13}, \citet{RocGeo16}, and \citet{WitTibHas09}. The first competitor was developed to estimate the rank of $\Lambda$, and the last three competitors were developed to estimate $\Lambda$. We used two versions of Ro{\v{c}}kov{\'a} and George's method. The first version used the expectation-maximization algorithm developed in \citet{RocGeo16}, and the second version added an extra step in every iteration of the algorithm that rotated the loadings matrix using the varimax criterion.

We evaluated the performance of the methods for estimating $\Lambda$ on simulated data using root mean square error, proportion of true positives, and proportion of false discoveries:
\begin{align}
  &\text{mean square error}  = \sum_{d=1}^p \sum_{j=1}^k ( |\lambda_{dj}^*| - |\hat \lambda_{dj}| )^2/ (pk),\quad \text{true positive rate} = |\hat \Mcal \cap \Mcal^* | /  |\Mcal^*| \nonumber\\  
  & \text{false discovery rate} = |\hat  \Mcal \backslash \Mcal^*| /  |\hat \Mcal|,
  \label{eqn:met}
\end{align}
where $\Lambda^*$ and $\hat \Lambda $ were the true and estimated loadings matrices and $\Mcal^*$ and $\hat \Mcal$ were the true and estimated locations of nonzero loadings. We assume that $\lambda_{dj}^*=0$ for any $d$ and $j=k^* + 1, \ldots, k$. Since $\lambda_{dj}^*$ and $\hat \lambda_{dj}$ could differ in sign, mean square error compared their magnitudes. 

\subsection{Simulated data analysis}
\label{sec:simulations}

The simulation settings were based on examples in \citet{KneSar11}. The number of dimensions varied as $p=50, 100, 250, 500, 2000$. The rank of every simulated loadings matrix was fixed at $k^*=5$. The magnitudes of nonzero loadings in a column were equal and decreased as $10$, $8$, $6$, $4$, and $2$ from the first to the fifth column. The signs of the nonzero loadings were chosen such that the columns of any loadings matrix were orthogonal, with a small fraction of overlapping nonzero loadings between adjacent columns: 
\begin{align*}
\lambda^*_{dj} =
  \begin{cases}
    2 (6 - j), \quad  &1 + (j -1 ) \tfrac{p}{k^*} \leq d \leq j \tfrac{p}{k^*},\quad 1 \leq j \leq k^*,\\
    -2 (6 - j), \quad  &1 + j \tfrac{p}{k^*} \leq d \leq (j + 1) \tfrac{p}{k^*},\quad  1 \leq j \leq k^*-1,\\
    -2 (6 - j), \quad  &(j - 1)  \tfrac{p}{k^*}\leq d \leq j \tfrac{p}{k^*} - 1,\quad  2 \leq j \leq k^*,\\
    0, \quad &\text{otherwise.}
  \end{cases}
\end{align*}
The error variances $\sigma^2_{d}$ increased linearly from $0$$\cdot$$01$ to $1$ for $d=1, \ldots, p$. Varying the  sample size as $n=50, 100, 250, 500, 5000$, data were simulated using model \eqref{eq:scal} for all combinations of $n$ and $p$. The simulation setup was replicated ten times and all five methods were applied in every replication by fixing the upper bound on the number of factors at $20$. The $\delta$-$\rho$ grid had dimensions $20 \times 20$ and $\log_{10} \delta$ increased linearly from $\log_{10} 2$  to $\log_{10} 10$ and $\log_{10} \rho$ increased linearly from $-3$  to 3 when $n > p$ and from $-2$ to $6$ when $n \leq p$. 

\begin{figure}[h]
  \begin{center}
    \includegraphics[scale=0.11]{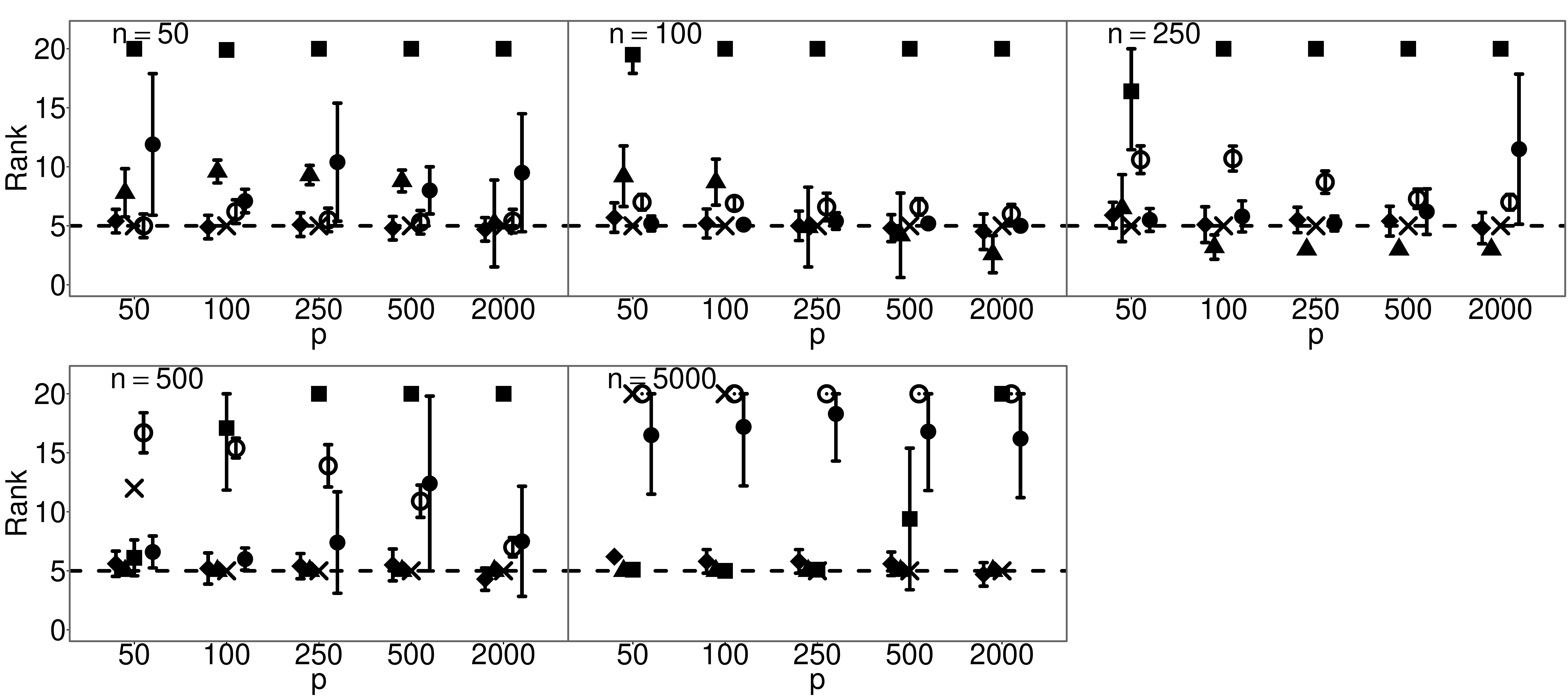}    
  \end{center}
  \caption{Rank estimate averaged across simulation replications for Caner \& Han's (times), Hirose \& Yamamoto's (solid square),  Ro{\v{c}}kov{\'a} and George's (circle), varimax version of Ro{\v{c}}kov{\'a} and George's (solid circle), and Witten et al.'s (solid diamond) methods, and our estimation algorithm (solid triangle). The horizontal line (dashed) represents the true number of factors. Error bars denote Monte Carlo errors.
 }
  \label{fig:rnks}
\end{figure}

All five methods had the same computational complexity of $O(p \log p)$ for one iteration, but their run-time differed depending on their implementations, with Witten et al.'s method being the fastest. Figure \ref{fig:rnks} shows that Hirose \& Yamamoto's  and both versions of Ro{\v{c}}kov{\'a} and George's method significantly overestimated $k^*$ for large $p$. Witten et al.'s method  slightly overestimated $k^*$ across all settings. Caner and Han's method showed excellent performance and accurately estimated $k^*$ across all simulation settings, except when $n=5000$ and $p=50, 100$. When $n$ was larger than 500, Assumption A.4 was satisfied and our method accurately estimated $k^*$ as 5 in every setting, performing better than Caner and Han's method when $n=5000$.

\begin{figure}[h]
  \begin{center}
    \includegraphics[scale=0.11]{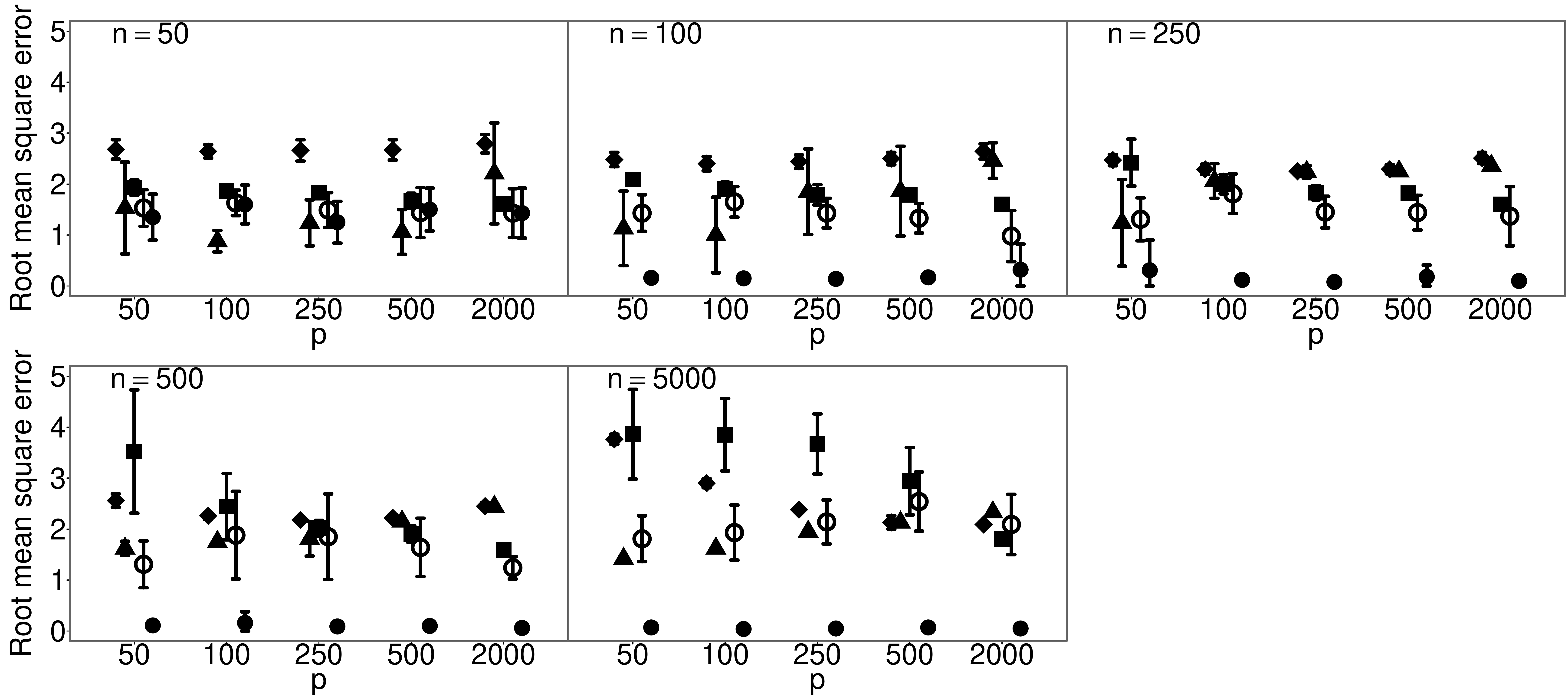}    
  \end{center}
  \caption{Root mean square error averaged across simulation replications for Hirose \& Yamamoto's (solid square),  Ro{\v{c}}kov{\'a} and George's (circle), varimax version of Ro{\v{c}}kov{\'a} and George's (solid circle), and Witten et al.'s (solid diamond) methods, and our estimation algorithm (solid triangle). Error bars denote Monte Carlo errors. }
  \label{fig:mse}
\end{figure}

\begin{figure}[h]
  \begin{center}
  \includegraphics[scale=0.11]{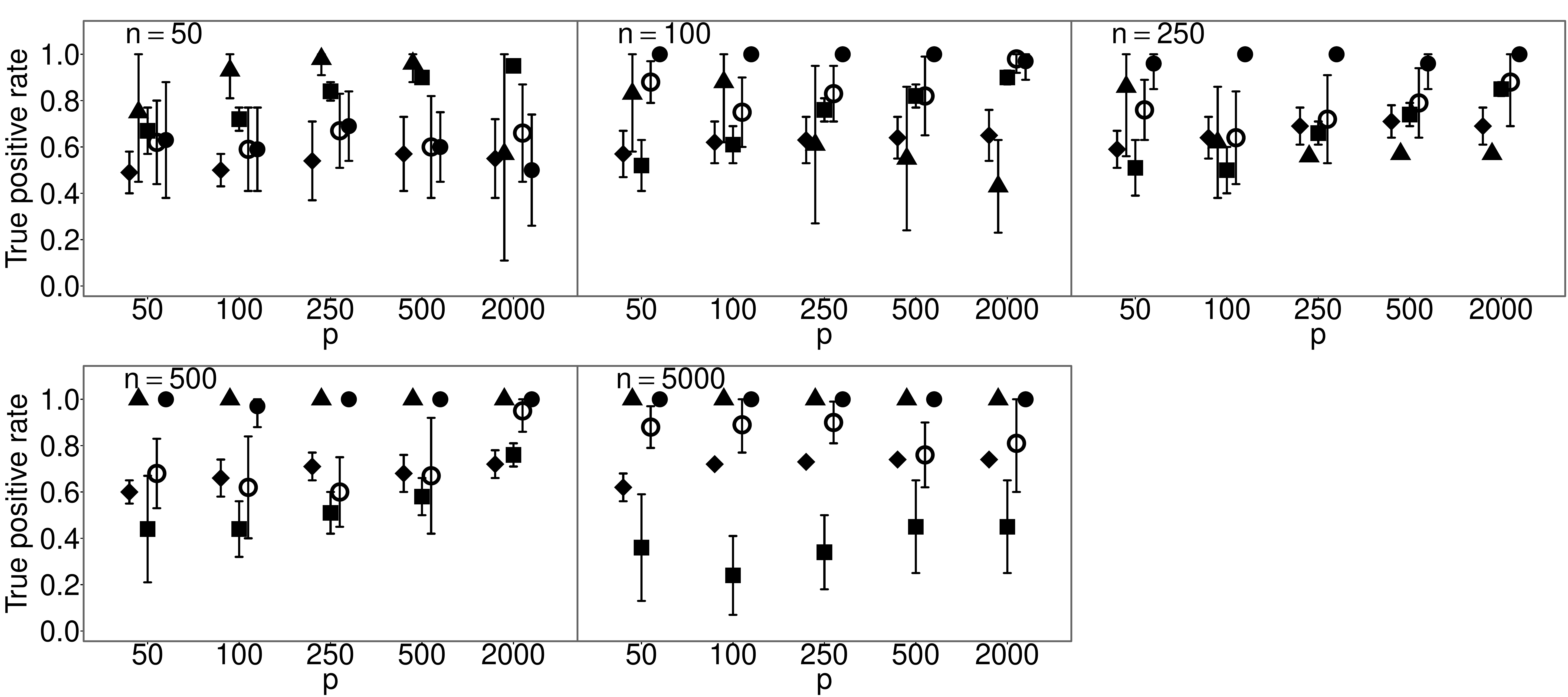}    
  \end{center}
  \caption{True positive rate averaged across simulation replications for Hirose \& Yamamoto's (solid square),  Ro{\v{c}}kov{\'a} and George's (circle), varimax version of Ro{\v{c}}kov{\'a} and George's (solid circle), and Witten et al.'s (solid diamond) methods, and our estimation algorithm (solid triangle). Error bars denote Monte Carlo errors. }
  \label{fig:tpr}
\end{figure}

\begin{figure}[h]
  \begin{center}
  \includegraphics[scale=0.11]{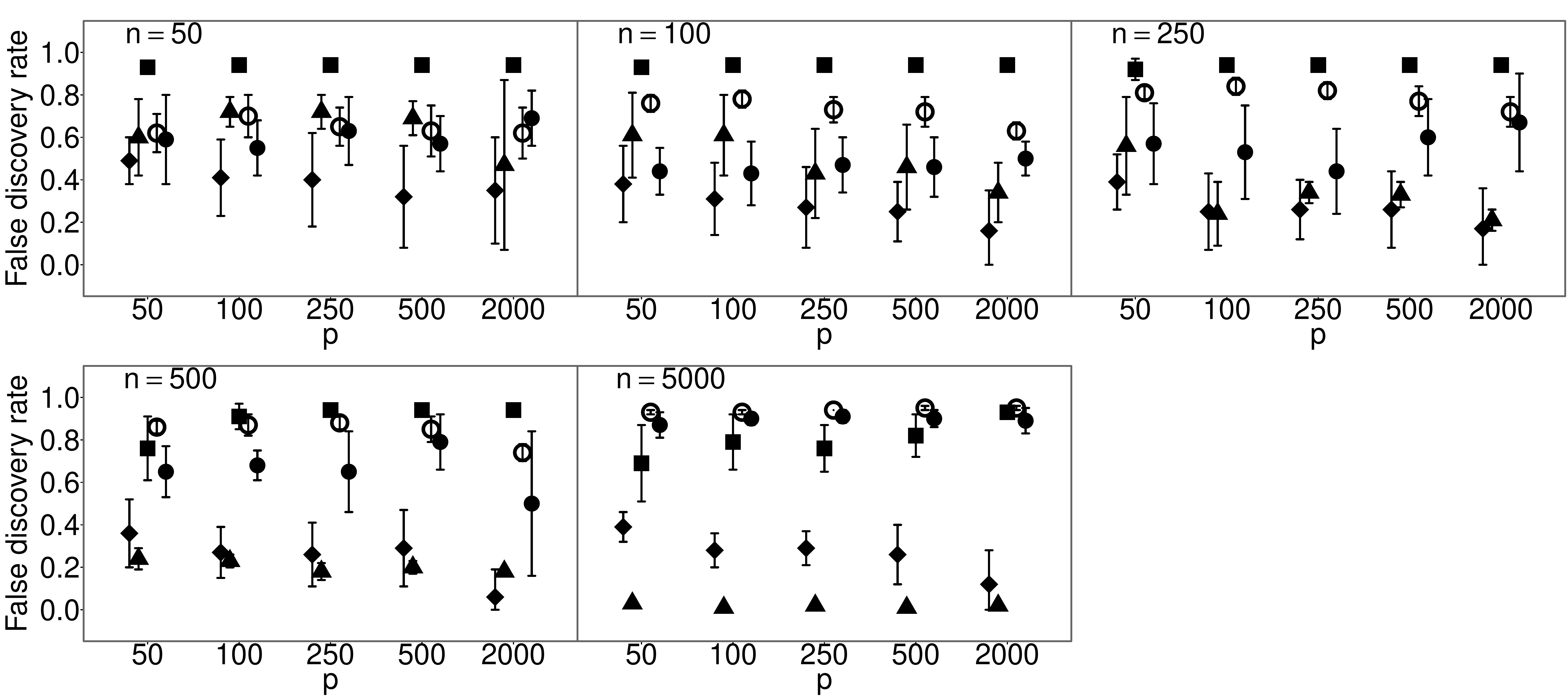}    
  \end{center}
  \caption{False discovery rate averaged across simulation replications for Hirose \& Yamamoto's (solid square),  Ro{\v{c}}kov{\'a} and George's (circle), varimax version of Ro{\v{c}}kov{\'a} and George's (solid circle), and Witten et al.'s (solid diamond) methods, and our estimation algorithm (solid triangle). Error bars denote Monte Carlo errors. }
  \label{fig:fdr}
\end{figure}

The four methods for estimating $\Lambda$ differed significantly in their root mean square errors, true positive rates, and false discovery rates; see Figures \ref{fig:mse}, \ref{fig:tpr}, and \ref{fig:fdr}. Hirose \& Yamamoto's method had the highest false discovery rates and the lowest true positive rates across most settings.  
Both versions of Ro{\v{c}}kov{\'a} and George's method estimated an overly dense $\Lambda$ across most settings, resulting in high true positive rates and high false discovery rates. The extra rotation step in the second version of Ro{\v{c}}kov{\'a} and George's method resulted in excellent mean square error performance; however, varimax rotation is a post-processing step. A similar step to reduce the mean square error could be added to our method; for example, by including a step to rotate the $\Lambda^0$ in step 3 of Algorithm \ref{xfa-alg} using the varimax criterion. When $n$ and $p$ were small, Witten et al.'s method achieved the lowest false discovery rates and our method achieved the highest true positive rates. When $n$ and $p$ were larger than 250 and 100, respectively, then Assumption A.4 was satisfied and our method simultaneously achieved the highest true positive rates and lowest false discovery rates while maintaining competitive mean square errors relative to the rotation-free methods.

\subsection{Microarray data analysis}
\label{sec:real-data-analysis}

We used gene expression data capturing aging in mice from the AGEMAP database \citep{Zahetal07}. There were 40 mice aged 1, 6, 16, and 24 months in this study. Each age group included 5 male and 5 female mice. Tissue samples were collected from 16 different tissues, including cerebrum and cerebellum, for every mouse. Gene expression levels in every tissue sample were measured on a microarray platform. After normalization and removing missing data, gene expression data were available for all $8,932$ probes across $618$ microarrays. We used a factor model to estimate the effect of latent biological processes on gene expression variation. 

AGEMAP data were centered before analysis following \citet{PerOwe10}. Gene expression measurements were represented by $Y \in \Re^{n \times p}$, where $n = 618$ and $p = 8932$. Further, age$_i$ represented the age of mouse $i$ and gender$_i$ was 1 if mouse $i$ was female and was 0 otherwise. Least square estimates of the intercept, age effect, and gender effect in the linear model $y_{id} = \beta_{0d} + \beta_{1d} \, \text{age}_{i} + \beta_{2d} \, \text{gender}_{i} + e_{id}$ ($i=1, \ldots, n$), with idiosyncratic error $e_{id}$,  were represented as $\hat \beta_{0d}$, $\hat \beta_{1d}$, and $\hat \beta_{2d}$. Using these estimates for $d=1, \ldots, p$, the mean-centered data were defined as
\begin{align*}
  \hat y_{id} = y_{id} - \hat \beta_{0d} + \hat \beta_{1d} \, \text{age}_{i} + \hat \beta_{2d} \, \text{gender}_{i}, \quad (i=1, \ldots, n; \, d=1, \ldots, p).
\end{align*}
Four mice were randomly held out, and all tissue samples for these mice in $\hat Y$ were used as test data. The remaining samples were used as training data. This setup was replicated ten times. All four methods were applied to the training data in every replication by fixing the upper bound on the number of factors at $10$. The $\delta$-$\rho$ grid had dimensions $20 \times 20$ and $\log_{10} \delta$ increased linearly from $\log_{10} 2$ to $\log_{10} 10$, and $\log_{10} \rho$ increased linearly from $-3$ to 6. 

The results for all five methods were stable across all ten folds of cross-validation. Caner and Han's, Hirose and Yamamoto's, both versions of Ro{\v{c}}kov{\'a} and George's, Witten et al.'s, and our method selected 10, 10, 10, 4, and 1, respectively, as the number of latent biological process $k^*$ across all folds. Our result matched the result in \citet{PerOwe10}, who confirmed the presence of one latent variable using rotation tests. Our simulation results and the findings in \citet{PerOwe10} strongly suggest that our method accurately estimated $k^*$ and the alternative methods overestimated $k^*$. 

We also estimated the factors for the test data. With $\hat y_i$ denoting test datum $i$ and $UDV^T$ denoting the singular value decomposition of $\Lambda$, the factor estimate of test datum $i$ was $n_T^{-1/2} U^T \hat y_i$, where $n_T$ was the number of samples in training data. \citet{PerOwe10} found that factor estimates for the tissue samples from cerebrum and cerebellum, respectively, had bimodal densities. We used the density function in R with default settings to obtain kernel density estimates of the factors. Hirose and Yamamoto's and both version of Ro{\v{c}}kov{\'a} and George's method estimated the number of factors as 10, which made the results challenging to interpret. Witten et al.'s method recovered bimodal densities in all four factors for both tissue samples, but it was unclear which of these four factors corresponded to the factor estimated by \citet{PerOwe10}. Our method estimated the number of factors as 1 and recovered the bimodal density in both tissue samples.



\section*{Acknowledgement}
This work was supported by grants from the National Institute of Environmental Health Sciences, the National Institutes of Health, and the National Science Foundation. The code used in the experiments is available at \url{https://github.com/blayes/xfa}.

\appendix

\section{Assumptions}

Assumptions A.0--A.4 follow from the theoretical setup for high-dimensional factor models in \cite{KneSar11}.  Assumption A.5 is based on results in \citet{ZouLi08} for variable selection.
\begin{itemize}
\item [A.0] Let $y_i =  w_i + e_i$, $E(y_i) = 0$, $\text{var}(y_i) = \Omega^*$,  $E(w_i) = 0$, $\text{var}(w_i) = \Lambda^* \Lambda^{*^T}$, $E(e_i) = 0$, $\text{var}(e_i) = \Sigma^*$ ($i = 1, \ldots, n$).
\item [A.1] There exist finite positive constants $D_0$, $D_3$,  $D_1 \leq D_2$ such that $E(y^2_{id}) \leq D_0$, $E(e_{id}^4) \leq D_3$, and $0 < D_1 \leq (\Sigma^*)_{dd} \leq D_2$
  ($i = 1, \ldots, n$; $d = 1, \ldots, p$).
\item [A.2] There exists a constant $C_0 \in (8, \infty)$ such that $\sum_{i=1}^n w_{ij} w_{il}/n$, $\sum_{i=1}^n e_{ij} e_{il}/n$, $\sum_{i=1}^n w_{ij} e_{il}/n$, $\sum_{i=1}^n y_{ij} y_{il}/n$ are $(C_0/n)$-subgaussian for every $j, l$ $\in$ $\{1, \ldots, p\}$. A random variable $X$ is $c$-subgaussian if $\text{pr}\{|X - E(X)| > t\} \leq  2 e^{-t^2 / (2c)}$ for any $t > 0$.
\item [A.3] Let $b_1 > \cdots > b_{k^*} > 0$ be the eigenvalues of $\Lambda^* \Lambda^{*^T}$, then there exists a $v_0$ such that $0 < v_0 \leq 1$, $pv_0 > 6 D_2$, $\underset{j, l \leq k^*, j \neq l}{\min}| b_j / p - b_l / p| \geq v_0$, and $\underset{j \leq k^*}{\min} \, b_j /p \geq v_0$.
\item [A.4] The sample size $n$ and dimension $p \geq n$ are large enough such that $C_0 (\log p / n)^{1/2} \geq D_0 / p$ and $v_0 \geq 6 \{D_2 /p + C_0 (\log p / n)^{1/2}\}$.
\item [A.5] Let $k$ be the upper bound on $k^*$ and $\delta$, $\rho$, $\alpha_j(\delta)$, and $\eta_j(\rho)$ ($j=1, \ldots, k$) are defined as in Lemma \ref{lem:gdp-2}. Then, $k = O(\log p)$, $\alpha_{j}(\delta) \rightarrow \infty$, ${n^{-1/2}\alpha_{j}(\delta)} \rightarrow 0$, and $(np)^{1/2}\eta_{j}(\rho) \rightarrow c_j > 0$ ($j=1, \ldots, k$) as $n \rightarrow \infty$, $n \leq p \rightarrow \infty$, and $\log p / n \rightarrow 0$.
\item [A.6] The elements of the set $\Mcal^*$ are fixed and do not change as $n$ or $p$ increase to $\infty$. 
\end{itemize}
Model \eqref{mdl-1} is recovered by substituting $w_i = \Lambda^* z_i$ in A.0. Assumption A.1 ensures that $\Omega^*$ is positive definite. Assumption A.2 ensures the empirical covariances are good approximations of the true covariances. Specifically, for any $t > 0$, 
  \begin{align*}
    \underset{1 \leq j, l \leq p}{\sup} \left| \frac{1}{n}\sum_{i=1}^n w_{ij} w_{il} - \text{cov}(w_{ij}, w_{il})\right| \leq t, \quad
    \underset{1 \leq j, l \leq p}{\sup} \left| \frac{1}{n}\sum_{i=1}^n e_{ij} e_{il} - \text{cov}(e_{ij}, e_{il})\right| \leq t,  \\
    \underset{1 \leq j, l \leq p}{\sup} \left| \frac{1}{n}\sum_{i=1}^n w_{ij} e_{il} \right| \leq t,  \quad
    \underset{1 \leq j, l \leq p}{\sup} \left| \frac{1}{n}\sum_{i=1}^n y_{ij} y_{il} - \text{cov}(y_{ij}, y_{il})\right| \leq t,
  \end{align*}
hold simultaneously with probability at least $A_t(n,p) = 1 - 8 p^2 e^{-nt^2 / (2C_0)}$. If $t_0 = C_0 (\log p / n)^{1/2}$, then $A_{t_0}(n,p) \rightarrow 1$ as $n$, $p \rightarrow \infty$ and $\log p / n \rightarrow 0$. Assumption A.3 guarantees identifiability of $\Lambda^{0}$ when $p$ is large and $v_0 \gg 1/p$. Assumption A.4 is required to ensure that $p^{-1/2} (\lambdah_d)^{1/2}\psih_{dj}$ is a root-$n$ consistent estimator of $p^{-1/2}{\lambda_{dj}}$ as $n \rightarrow \infty$, $n \leq p \rightarrow \infty$, and $\log p / n \rightarrow 0$. 

One additional assumption is required to relate $\textsc{ebic}_{\gamma}(\Mcal)$ and $\pi_{\Mcal} = \text{pr}(\Mcal \mid Y)$,
\begin{itemize}
\item [A.7] $p = O(n^{\kappa})$ for a fixed constant $\kappa \geq 1$ such that $\gamma > 1 - 1 / (2 \kappa)$. 
\end{itemize}
Assumption A.7 and Equation 4.6 in Theorem 3 of \cite{KneSar11} imply that $\lambdah_l > 0$ for any $l$ such that $1 \leq l \leq 2k < p$ because $(\log p)^{3/2} / n^{1/2} \rightarrow 0$ as $n \rightarrow \infty$.

\bibliographystyle{Chicago}
\bibliography{papers}

\clearpage

\renewcommand\thesection{\arabic{section}}
\renewcommand\thesubsection{\thesection.\arabic{subsection}}
\renewcommand\thesubsubsection{\thesubsection.\arabic{subsubsection}}

\setcounter{section}{0}

\begin{center}
\textbf{\Large Supplementary Material for Expandable Factor Analysis}
\end{center}

\section{Expectation-maximization algorithm for expandable factor analysis}
\subsection{Estimation of $\Lambda$ and $\Sigma$}
\label{sec:log-posterior-based}

Define the following quantities using mean-centered data:
\begin{align*}
  &S_{y y} = \frac{1}{n} \sum_{i=1}^n y_{i} y_{i}^T, \quad S_{zz} = \frac{1}{n} \sum_{i=1}^n z_{i} z_{i}^T, \quad S_{yz} = \frac{1}{n} \sum_{i=1}^n y_{i} z_{i}^T, \quad \Omega = \Lambda \Lambda^T + \Sigma,\\
  &\Delta = I_k - \Lambda^T \Omega^{-1} \Lambda, \quad G = \Omega^{-1} \Lambda, \quad F =  \Delta + G^T S_{y y} G, \quad L = S_{y y}G,
\end{align*}
where $I_k$ is the $k \times k$ identity matrix. We place Jeffreys' prior on the error variances, $\pi(\sigma_{d}) \propto \sigma^{-1}_{d}$ ($d=1, \ldots, p$). Let $\Lambda^{(t)}$ and $\Sigma^{(t)}$ be the estimates of $\Lambda$ and $\Sigma$ at iteration $t$, then the conditional expectations of $S_{zz}$, $S_{yz}$, and complete data log likelihood at iteration $(t+1)$ are 
\begin{align}
  \label{eq:em1}
  E \big( S_{z z} \mid Y, \Lambda^{(t)}, \Sigma^{(t)} \big) &= \Delta^{(t)} + G^{(t)^T} S_{y y} G^{(t)} = F^{(t)}, \quad  E \big(  S_{y z} \mid Y, \Lambda^{(t)}, \Sigma^{(t)} \big) =  L^{(t)}, \nonumber\\
 Q(\Lambda, \Sigma \mid \Lambda^{(t)}, \Sigma^{(t)}) &= E \big\{ (npk)^{-1} \log p(Z, \Lambda, \Sigma \mid Y, \Lambda^{(t)}, \Sigma^{(t)}, \alpha_{1:k}, \eta_{1:k}) \big\} \nonumber \\
                                                            &= - \sum_{d = 1}^p \bigg[\frac{1}{2pk}  \frac{\big(S_{y y} \big)_{dd} + \big\{ \Lambda E \big( S_{z z} \mid Y, \Lambda^{(t)}, \Sigma^{(t)} \big) \Lambda^{T} \big\}_{dd}}{\sigma_{d}^{2}} - \nonumber \\
                                                            &\qquad \qquad \;  \frac{1}{2pk} \frac{2 \big\{ E\big( S_{y z} \mid Y, \Lambda^{(t)}, \Sigma^{(t)} \big) \Lambda^{T} \big\}_{dd}}{\sigma_{d}^{2}} \bigg] \nonumber \\
  &\quad - \sum_{d=1}^{p} \sum_{j=1}^{k} \frac{\alpha_j + 1}{npk} \log \bigg(1 + \frac{|\lambda_{dj}|}{\eta_j} \bigg) - \frac{n+2} {2npk} \sum_{d=1}^p \log \sigma_{d}^{2} \nonumber\\
  &\equiv - \sum_{d=1}^p \log p_{\mathrm{mis}} (\lambda_d, \sigma_{d}^{2} \mid S_{y y}, F^{(t)}, L^{(t)}) - \frac{n+2} {2npk} \sum_{d=1}^P \log \sigma_{d}^{2},
\end{align}
where the superscript $(t)$ denotes the dependence on $\Lambda^{(t)}$ and $\Sigma^{(t)}$. The objective (\ref{eq:em1}) splits into $p$ separate terms, and term $d$ depends on $\lambda_d$ and $\sigma_{d}^{2}$; therefore, \eqref{eq:em1} is maximized by repeating the following two until steps until convergence to a fixed point:
\begin{enumerate}
\item  For $d = 1, \ldots, p$,
  \begin{enumerate}
  \item fix $\sigma_{d}^{2}$ at $\sigma_{d}^{2^{(t)}}$ in
    \begin{align}
  \log p_{\mathrm{mis}} (\lambda_d, \sigma_{d}^{2} \mid S_{y y}, F^{(t)}, L^{(t)})  + \frac{n+2} {2npk} \log \sigma_{d}^{2}, \label{pen-reg-p}
\end{align}
and minimize with respect to $\lambda_d$ to estimate $\lambda_d^{(t+1)}$;
\item fix $\lambda_d$ at $\lambda_d^{(t+1)}$ in \eqref{pen-reg-p} and minimize  \eqref{pen-reg-p} with respect to $\sigma_{d}^{2}$ to estimate $\sigma_{d}^{2^{(t+1)}}$.
  \end{enumerate}
\item Increment $t$ to $(t+1)$.
\end{enumerate}

\subsection{Block coordinate descent algorithm for estimation of $\Lambda$}
\label{sec:block-coord-desc}

We use local linear approximation of the objective \eqref{pen-reg-p} to derive a new block coordinate descent algorithm. We suppress the superscript $(t)$ in $w_d$ and $X$ to ease notation. The algorithm initializes  $\widetilde \Lambda^0$ at $\Lambda^{\mathrm{lla}^{(t)}}$ and updates $\widetilde \Lambda_{dj}^{{(i)}}$ using \eqref{pen-reg-p} as
\begin{align}
  \widetilde \lambda_{dj}^{{(i+1)}} = \underset{\lambda_{dj}}{\argmin} \, \frac{\lambda_{dj}^2 X^T_j X_j + 2 \lambda_{dj} (\widetilde \Lambda_{d,(-j)}^{{(i)^T}}  X^T_{(-j)} X_j - X_j^T w_{d})}{2} + \frac{(\alpha_j + 1) \sigma_{d}^{2^{(t)}}}{(\eta_j + | \lambda_{dj}^{(t)}|) n} |\lambda_{dj}| \nonumber 
\end{align}
successively for $j=1, \ldots, k$ in the $(i+1)$th cycle. This objective function is convex and its optimum is  
\begin{align}
  \widetilde \lambda_{dj}^{{(i+1)}} = \frac{\sgn(s_{dj}^{{(i)}})}{f_{jj}} \left( | s_{dj}^{(i)} | -  \frac{(\alpha_j + 1) \sigma_{d}^{2^{(t)}}}{(\eta_j + | \lambda_{dj}^{(t)}|) n}\right)_{+}, \label{lam-st}
\end{align}
where $s_{dj}^{(i)} = X_j^T w_{d} - \widetilde \Lambda_{d,(-j)}^{(i)}  X^T_{(-j)} X_j$ and $f_{jj} = X^T_j X_j$. We also exploit the form of (\ref{lam-st}) and use it to update the $k$th column of $\widetilde \Lambda^{{(i)}}$. This leads to $k$ block updates for $\widetilde \Lambda^{{(i)}}$ in a single cycle of the coordinate descent algorithm. These updates are repeated until the change in $\widetilde \Lambda$ is negligible. We then set $\Lambda^{\mathrm{lla}^{(t+1)}} = \widetilde \Lambda^{(\infty)}$. We have implemented this algorithm in R \citep{R16} using the glmnet package \citep{Frietal10}.

\subsection{Root-$n$ consistent estimates of $\Lambda$ and $\Sigma$}

Let $S_{yy}$ be the empirical covariance matrix of mean-centered data and $\lambdah_d$ and $ \widehat \psi_d$ ($d = 1, \ldots, p$) be its eigenvalues and eigenvectors, then 
\begin{align}
  S_{yy} = Y^TY/n = \sum_{d=1}^p \lambdah_d \widehat \psi_d \widehat \psi_d^T \label{eig-dec1}
\end{align}
is the eigen decomposition of $S_{yy}$. Use \eqref{eig-dec1} to define
\begin{align*}
  \lambda^0_{dj} = \lambdah_j^{1/2} \widehat \psi_{dj} \quad (d = 1, \ldots, p; \; j = 1, \ldots, k).
\end{align*}
An application of Theorem 2 in \citet{KneSar11} shows that ${\lambda_{dj}^0} / p^{1/2}$ is a root-$n$ consistent estimator of ${\lambda_{dj}} /p^{1/2}$ when $n \leq p$. Equations 4.3 and 4.4 in \citet{KneSar11} and Assumptions A.1--A.4 in the main paper imply that there exist universal positive constants $D_0, D_1,$ and $C_0$ such that
\begin{align*}
   \frac{\lambda_{dj}^2} {p}  \leq \frac{D_0 - D_1}{p}, \quad \frac{\lambda_{dj}^{0^2}} {p} \leq \frac{D_0 + C_0 \left( {\log p} / {n} \right)^{1/2}}{p}
\end{align*}
with probability at least $A(n, p) = 1 - 8 p^{2 - C_0 / 2} \rightarrow 1$ as $n \rightarrow \infty$, $n \leq p \rightarrow \infty$, and ${\log p}/{n} \rightarrow 0$. This implies that 
\begin{align}
  \left|\frac{\lambda_{dj}} {p^{1/2}} - \frac{\lambda_{dj}^0} {p^{1/2}} \right| \leq \left( \frac{D_0 - D_1}{p} \right)^{1/2} + \left\{ \frac{D_0 + C_0 \left( {\log p}/{n} \right)^{1/2}}{p} \right\}^{1/2} \label{rootn-1}
\end{align}
with probability at least $A(n, p)$. Since ${\log p}/{n} \rightarrow 0$, ${\log p}/{n} \leq D_0^2 / C_0^2$ for large $n$ and $p$ and \eqref{rootn-1} reduces to
\begin{align*}
  \left| \frac{\lambda_{dj}^0} {p^{1/2}}  -  \frac{\lambda_{dj}} {p^{1/2}} \right| \leq \left( \frac{2D_0}{p} \right)^{1/2} \leq \left( \frac{2D_0}{n} \right)^{1/2}
\end{align*}
with probability at least $A(n, p) \rightarrow 1$ as $n \rightarrow \infty$, $n \leq p \rightarrow \infty$, and ${\log p}/{n} \rightarrow 0$. This shows that ${\lambda^0_{dj}}/{p^{1/2}}$ is a root-$n$ consistent estimator of ${\lambda_{dj}} / {p^{1/2}}$. Theorem 3 in \citet{KneSar11} implies that $\sigma^{2^0}_{d} = (S_{ y y} - \Lambda^0 \Lambda^{0^T})_{dd}$ is a root-$n$ consistent estimator of $\sigma^{2}_{d}$ for overfitted factor models.

We also prove a result that is used in the proof for asymptotic normality of nonzero loadings.
\begin{lemma}\label{lem:bdd-var}
  If Assumptions A.0--A.4 in the main paper hold, then $E(\lambda^{0^2}_{dj}) < \infty$ ($d=1, \ldots, p$; $j=1, \ldots, k$).
\end{lemma}
\begin{proof}
  Using \eqref{eig-dec1},
  \begin{align}
    \label{eq:lem1sup}
    E(\lambda^{0^2}_{dj}) = E(\lambdah_j^2 \widehat \psi_{dj}^2) \overset{(i)}{\leq} E(\lambdah_j^2) = \int_{0}^{\infty} \text{pr}(\lambdah_j^2 > t) \, dt \leq (D_2 + D_0)^2 + \int_{(D_2 + D_0)^2}^{\infty} \text{pr}(\lambdah_j^2 > t) \, dt,
  \end{align}
  where $(i)$ follows because $\sum_{d=1}^p \widehat \psi_{dj}^2 = 1$. Equation 4.1 of Theorem 2 in \citet{KneSar11} implies that for some $\zeta_j \geq 0$, 
  \begin{align*}
    8 / p^{C_0/2 - 2} &\geq \text{pr} \left\{ |\lambdah_j / p - \zeta_j / p| >  D_2 / p + C_0 (\log p / n)^{1/2}  \right\} \\
                   &\overset{(ii)}{\geq} \text{pr} \left\{ |\lambdah_j / p - \zeta_j / p| >  (C_0 D_2 / D_0 + C_0) (\log p / n)^{1/2}  \right\} \\    
                   &\geq \text{pr} \left\{ \lambdah_j / p >   \zeta_j / p +  (C_0 D_2 / D_0 + C_0) (\log p / n)^{1/2}  \right\} \\    
                   &\geq \text{pr} \left\{ \lambdah_j / p >   (C_0 D_2 / D_0 + C_0) (\log p / n)^{1/2}  \right\} \\
                   &= \text{pr} \left\{ \lambdah_j^2 >   (C_0 D_2 / D_0 + C_0)^2 p^2 \log p / n \right\},
  \end{align*}
where $(ii)$ follows because $C_0 (\log p /n)^{1/2} > D_0 / p$ by Assumption A.4 in the main paper. Substituting $t = (C_0 D_2 / D_0 + C_0)^2 p^2 \log p / n$ in \eqref{eq:lem1sup} implies that
\begin{align*}
  \text{pr} (\lambdah_j^2 > t) \leq 8 (C_0 D_2 / D_0 + C_0)^{C_0/2 - 2} (\log p / n)^{C_0 / 2 - 2} t^{1 - C_0 / 4}, \quad t \geq (D_0 + D_2)^2.
\end{align*}
Therefore, $\int_{(D_2 + D_0)^2}^{\infty} \text{pr}(\lambdah_j^2 > t) \, dt < \infty$ for $C_0 \in (8, \infty)$, which in turn shows that $E(\lambda^{0^2}_{dj})$ is bounded because $\log p / n \goesto 0$.

\end{proof}

\subsection{Computational complexity}
\label{sec:comp-compl-impl}
The computational complexity of the estimation algorithm equals the cost of performing $p$ penalized regression problems of dimension $k = O(\log p)$. Our estimation algorithm requires $O(np^2 + p \log^2 p)$ time upfront to calculate $S_{yy}$ and its eigen decomposition. Estimation of $G, \Delta, F$, and $L$ in (\ref{eq:em1}) involves $k$-dimensional matrix multiplications and inversions of $O(\log^3 p)$ time complexity. Using these matrices, one iteration of the block coordinate descent algorithm has $O(\log p)$ time complexity for dimension $d$ ($d=1, \ldots, p$). The total time complexity of each iteration is $O(p \log p + \log^3 p)$; therefore, the time complexity of $T$ iterations of the expectation-maximization algorithm is $O(T p \log p)$.

\section{Properties of the multiscale generalized double Pareto prior}

\subsection{Proof of Lemma 1}
\label{sec:proof-lemma-1}

If $\Thetabb$ is the support of multiscale generalized double Pareto prior on $\Lambda$, then 
\begin{align*}
  \text{pr}(\Thetabb) = &\text{pr} \left( \Lambda \mid \underset{1 \leq d \leq p}{\max} \sum_{k=1}^{\infty} \lambda_{dk}^2 < \infty \right) \geq 1 - \lim_{t \uparrow \infty} \sum_{d=1}^p \text{pr} \left( \Lambda \mid \sum_{k=1}^{\infty} \lambda_{dk}^2 \geq t \right) \geq 1 - p \lim_{t \uparrow \infty} \frac{\sum_{k=1}^{\infty} V(\lambda_{1k})}{t}. 
\end{align*}
Since $\lambda_{1k}$ follows generalized double Pareto distribution with parameters $(\alpha_k, \eta_k)$,  $V(\lambda_{1k}) = 2 \eta^2_k (\alpha_k -1)^{-1}  (\alpha_k -2)^{-1}$ for $\alpha_k > 2$ and
\begin{align}
  \label{eq:lem1eq}
  \sum_{k=1}^{\infty} V(\lambda_{1k}) & \leq 2 \sum_{k=1}^{\infty} \frac{\eta_k^2}{\alpha^{2}_k} \bigg(1 - \frac{2}{\alpha_k} \bigg)^{-2} \leq \{2 + O(1)\} \sum_{k=1}^{\infty} \frac{\eta^2_k}{\alpha^{2}_k}. 
\end{align}
This summation is finite if $\alpha_k > 2$ and ${\eta_k}/{\alpha_k} = O (k^{-m})$ for $m>0$$\cdot$$5$; therefore, pr$(\Thetabb) = 1$.

\subsection{Proof of  Lemma 2}
\label{sec:proof-lemma-2}
Let $ k(p, \delta, \epsilon)$ be such that $\text{pr}\{\Omega^{ k} \mid d_{\infty}(\Omega, \Omega^{ k}) \geq \epsilon\} \leq \epsilon$ for any $\epsilon > 0$. Then,
\begin{align*}
  \text{pr}\{d_{\infty}(\Omega, \Omega^{ k}) \geq \epsilon\} \overset{(i)}{\leq} \sum_{i=1}^{p} \sum_{j=1}^{p} \text{pr} ( | \Omega_{ij} - \Omega_{ij}^{ k} | \leq \epsilon) \overset{(ii)}{\leq} \frac{p^2}{\epsilon} \sum_{l= k + 1}^{\infty} E(|\lambda_{1l}|^2),
\end{align*}
where $(i)$ follows from the union bound and $(ii)$ follows from Markov's inequality and the independence of $\lambda_{ik}$s. The assumptions in Lemma 2 of the main paper and \eqref{eq:lem1eq} imply that
\begin{align*}
  \frac{p^2}{\epsilon} \sum_{l= k + 1}^{\infty} E(|\lambda_{1l}|^2) = \text{constant} \, \frac{p^2}{\epsilon} \delta^{-2  k} \leq \epsilon \implies k =  O \left(\log^{-1} \delta \log \tfrac{p}{\epsilon} \right).
\end{align*}

\section{Theoretical properties of $\Lambda^{\mathrm{lla}}$ and $\Sigma^{\mathrm{lla}}$}

\subsection{Proof of Theorem 1}
\label{sec:proof-theorem-1}

Let $\theta  = (\Lambda, \Sigma)$. Then, the objective function in \eqref{eq:em1} is
\begin{align}
  \Lcal(\theta) = \Lcal_{\mathrm{ML}}(\theta)  - \sum_{d=1}^{p} \sum_{j=1}^{k} \frac{\alpha_j + 1}{npk} \log \bigg(1 + \frac{|\lambda_{dj}|} {\eta_j} \bigg) - \frac{n+2} {2npk} \sum_{d=1}^p \log \sigma_{d}^{2}, \label{em-2} 
\end{align}
where $\Lcal_{\mathrm{ML}}(\theta)$ is the log likelihood of $\theta$ scaled by $npk$. This leads to the $Q$-function
\begin{align}
  Q(\theta \mid \theta^{(t)}) = - &\sum_{d=1}^p \log p_{\mathrm{mis}} (\lambda_d, \sigma_{d}^{2} \mid S_{yy}, F^{(t)}, L^{(t)}) - \frac{n+2} {2npk} \sum_{d=1}^p \log \sigma_{d}^{2}. \label{q-fun-em}
\end{align}
The local linear approximation of (\ref{q-fun-em}) is
\begin{align}
  Q_{\mathrm{LLA}}(\theta \mid \theta^{(t)})  &= - \sum_{d = 1}^p \frac{\big(S_{y y} \big)_{dd} + \big(\Lambda F^{(t)} \Lambda^{T} \big)_{dd} - 2 \big(L^{(t)} \Lambda^{T} \big)_{dd}}{2pk\sigma_{d}^{2}} - \frac{n+2} {2npk} \sum_{d=1}^p \log \sigma_{d}^{2}  \nonumber \\
                                              &\quad - \sum_{d=1}^{p} \sum_{j=1}^{k} \frac{\alpha_j + 1}{npk} \left\{ \log \bigg(1 + \frac{|\lambda^{(t)}_{dj}|}{\eta_j} \bigg) + \frac{ \text{sign}(\lambda^{(t)}_{dj})}{\eta_j + |\lambda^{(t)}_{dj}|} (\lambda_{dj} - \lambda^{(t)}_{dj})  \right\} \nonumber \\
                                              &= Q_{\mathrm{ML}}(\theta \mid \theta^{(t)}) - \frac{n+2} {2npk}\sum_{d=1}^p \log \sigma_{d}^{2}  \nonumber \\
                                              &\quad - \sum_{d=1}^{p} \sum_{j=1}^{k} \frac{\alpha_j + 1}{npk} \left\{ \log \bigg(1 + \frac{|\lambda^{(t)}_{dj}|}{\eta_j} \bigg) + \frac{\text{sign}(\lambda^{(t)}_{dj})}{\eta_j + |\lambda^{(t)}_{dj}|}  (\lambda_{dj} - \lambda^{(t)}_{dj}) 
\right\}, \label{em-3}
\end{align}
where $Q_{\mathrm{ML}}(\theta \mid \theta^{(t)})$ is the $Q$-function that corresponds to $\Lcal_{\mathrm{ML}}(\theta)$. Theorem 1 of \citet{DemLaiRub77} shows that $Q_{\mathrm{ML}}(\theta^{(t)} \mid \theta^{(t)}) = \Lcal_{\mathrm{ML}}(\theta^{(t)})$, and using this in \eqref{em-2} and \eqref{em-3} shows that $Q(\theta^{(t)} \mid \theta^{(t)}) = \Lcal(\theta^{(t)})$ and $Q_{\mathrm{LLA}}(\theta^{(t)} \mid \theta^{(t)}) = \Lcal(\theta^{(t)})$. Subtracting \eqref{em-3} from \eqref{em-2} yields 
\begin{align}
\Lcal(\theta) - Q_{\mathrm{LLA}}(\theta \mid \theta^{(t)})  = 
\Lcal_{\mathrm{ML}}(\theta) - Q_{\mathrm{ML}}(\theta \mid \theta^{(t)}) + \sum_{d=1}^{p} \sum_{j=1}^{k}\frac{\alpha_j + 1}{npk} l_{dj}(\lambda_{dj} \mid \lambda^{(t)}_{dj}),
\end{align}
where
\begin{align}
  l_{dj}(\lambda_{dj} \mid \lambda^{(t)}_{dj}) = \log \bigg(1 + \frac{|\lambda^{(t)}_{dj}|}{\eta_j} \bigg) + \frac{\text{sign}(\lambda^{(t)}_{dj})}{\eta_j + |\lambda^{(t)}_{dj}|}  (\lambda_{dj} - \lambda^{(t)}_{dj}) - \log \bigg(1 + \frac{|\lambda_{dj}|} {\eta_j} \bigg).
\end{align}
The log function is concave and is majorized by its tangent, so $l_{dj}(\lambda_{dj} \mid \lambda^{(t)}_{dj}) \geq 0$ for any $|\lambda_{dj}| \geq 0$; therefore, $\Lcal(\theta) - Q_{\mathrm{LLA}}(\theta \mid \theta^{(t)}) \geq 0$ because $\Lcal_{\mathrm{ML}}(\theta) - Q_{\mathrm{ML}}(\theta \mid \theta^{(t)}) \geq 0$ using Lemma 1 and Theorem 1 in \citet{DemLaiRub77}. If $\theta^{(t+1)}$ maximizes $Q_{\mathrm{LLA}}(\theta \mid \theta^{(t)})$, then
\begin{align}
  \Lcal(\theta^{(t+1)}) \geq Q_{\mathrm{LLA}}(\theta^{(t+1)} \mid \theta^{(t)}) \geq  Q_{\mathrm{LLA}}(\theta^{(t)} \mid \theta^{(t)}) = \Lcal(\theta^{(t)}),
\end{align}
where the last equality follows from (\ref{em-3}). The objective \eqref{eq:em1} is bounded in probability on the parameter space, so the sequence $\{\Lcal(\theta^{(t)})\}_{t=1}^{\infty}$ converges to some $\Lcal(\theta^{(\infty)})$. Using Proposition 1 in \citet{ZouLi08}, $\theta^{(t)}$ converges to the stationary point $\theta^{(\infty)}$. 

\subsection{Proof of asymptotic normality of nonzero loadings and consistency of estimated $\Lambda$}
\label{sec:proof-theorem-2}

The proof has two steps. First, we show asymptotic normality of nonzero loadings. Second, we use results of the first step to show consistency of the estimated loadings. 

{Step 1.} Let $\lambda^0_{dj} / p^{1/2}$ and $\sigma_{d}^{2^0} $ are the root-$n$ consistent sequence of estimators of $\lambda^*_{dj} / p^{1/2}$ and $\sigma_{d}^{2^*} $ ($d=1, \ldots, p$; $j=1, \ldots, k$) as $n \rightarrow \infty$, $n \leq p \rightarrow \infty$, and $\log p /n \rightarrow 0$, then imputing $Z$ based on the eigen decomposition  of $Y^TY/n$ in \eqref{eig-dec1} implies that 
\begin{align}
  \hat \Lambda = \underset{\begin{subarray}{c}\lambda_{d}\\
  d=1, \ldots, p
  \end{subarray}} {\argmin} \, \sum_{d=1}^p \frac{\| y_d / p^{1/2} - Z^0 \lambda_{d} / p^{1/2} \|^2}{2\sigma_{d}^{2^0} / p}
  + \sum_{d=1}^{p} \sum_{j=1}^{k} \frac{\alpha_{j} + 1}{\eta_j + |\lambda_{dj}^0|/ p^{1/2}}  |\lambda_{dj}/ p^{1/2}|, \label{eqn-a1}
\end{align}
where $\hat \Lambda$ is the estimate of $\Lambda$ obtained using the estimation algorithm of expandable factor analysis, 
\begin{align}
  \label{eq:zmat}
  &\sigma^{2^0}_{d} = \sum_{d=k+1}^p \lambdah_d \widehat \psi_{dj}^2 ,\quad \lambda^0 = {\lambdah_{j}^{1/2}} \widehat \psi_{dj}, \quad (d=1, \ldots, p; \; j=1, \ldots, k), \quad  \nonumber \\
  &Z^0= Y  \left( \lambdah_1^{-1/2} \widehat \psi_1 , \ldots, \lambdah_k^{-1/2}\widehat \psi_k \right).
\end{align}
Again using \eqref{eig-dec1}, 
\begin{align}
  \label{eq:td1}
  \frac{Z^{0^T}Z^0}{n} =  \left( \lambdah_1^{-1/2} \widehat \psi_1, \ldots, \lambdah_k^{-1/2} \widehat \psi_k \right)^T  Y^TY/n  
  \left( \lambdah_1^{-1/2} \widehat \psi_1, \cdots, \lambdah_k^{-1/2} \widehat \psi_k \right)= I_k.
\end{align}
If $U$ is a $p \times k$ matrix independent of $n$ and $p$ and $u_d^T$ represents row $d$ of $U$, then define
\begin{align}
  V_n(U) = \sum_{d=1}^p \frac{\left\|\frac{y_d}{p^{1/2}} - Z^0 \left(\frac{\lambda_{d}^*}{p^{1/2}} + \frac{u_d}{({np})^{1/2}} \right) \right\|^2} {2\sigma_{d}^{2^0} / p}
  + \sum_{d=1}^{p} \sum_{j=1}^{k} \frac{\alpha_{j} + 1} {\eta_j + \frac{|\lambda_{dj}^0|} {p^{1/2}}} \, \left|\frac{\lambda^*_{dj}} {p^{1/2}}  + \frac{u_{dj}}{({np})^{1/2}} \right|
  , \label{eqn-a2}
\end{align}
where vectors are added component-wise. Substitute $u_{dj} = 0$ $(d=1, \ldots, p$; $j = 1, \ldots, k$) in (\ref{eqn-a2}) to obtain 
\begin{align}
  V_n(0) = \sum_{d=1}^p \frac{\left \|\frac{y_d}{p^{1/2}} - Z^0\frac{\lambda_{d}^*}{p^{1/2}} \right \|^2} {2\sigma_{d}^{2^0} / p}
  + \sum_{d=1}^{p} \sum_{j=1}^{k} \frac{\alpha_{j} + 1} {\eta_j + \frac{|\lambda_{dj}^0|} {p^{1/2}}} \, \left|\frac{\lambda^*_{dj}} {p^{1/2}}  \right|. \label{eqn-a3}
\end{align}
Using \eqref{eq:zmat} and \eqref{eq:td1}, 
\begin{align}
  V_n(U) - V_n(0) = &\sum_{d=1}^p \frac{u_d^T u_d}{2 \sigma_{d}^{2^0}}  -  \sum_{d=1}^p \frac{n^{1/2} u_d^T } { \sigma_{d}^{2^0}} \left( \frac{Z^{0^T} y_d}{n} - \lambda_d^*  \right)  + \nonumber\\
  & \sum_{d=1}^{p} \sum_{j=1}^{k} \frac{\alpha_{j} + 1} {\eta_j + \frac{|\lambda_{dj}^0|} {p^{1/2}}} \left(  \left|\frac{\lambda^*_{dj}} {p^{1/2}}  + \frac{u_{dj}}{({np})^{1/2}} \right| - \left|\frac{\lambda^*_{dj}} {p^{1/2}}  \right| \right) \nonumber\\
 \equiv &\sum_{d=1}^{p} T_{1d} - \sum_{d=1}^{p} T_{2d} + \sum_{d=1}^{p} \sum_{j=1}^k T_{3dj}. \label{eqn-a4}
\end{align}

The limiting forms of all the terms in \eqref{eqn-a4} are derived next. First, we obtain the limiting form of $T_{1d}$ in \eqref{eqn-a4}. Because $\sigma_{d}^{2^0}$ is a root-$n$ consistent estimator of $\sigma_{d}^{2^*}$, $T_{1d} \goesto  {(u_d^T u_d) }/ ({2\sigma^{2^*}_{d}}) $ in probability as $n \rightarrow \infty$, $n \leq p \rightarrow \infty$, and $\log p /n \rightarrow 0$ using Slutsky's theorem. Second, we obtain the limiting form of $T_{2d}$ in \eqref{eqn-a4}. Lemma \ref{lem:bdd-var} shows that variance of $\lambda_{dj}^0$ ($d=1, \ldots, p$; $j=1, \ldots, k$) is bounded, so using \eqref{eig-dec1}, Slutsky's theorem, and the central limit theorem, 
\begin{align}
  \label{eq:td2-4}
  T_{2d} = n^{1/2} \left( \lambda_{d1}^0 - \lambda_{d1}^*, \, \ldots, \, 
\lambda_{dk}^0 - \lambda_{dk}^* \right) \frac{u_d } { \sigma_{d}^{2^0}} \goesto  \frac{u_d^T r_d} { \sigma_{d}^{2^*}} \quad (d=1, \ldots, p)
\end{align}
as $n \rightarrow \infty$, $n \leq p \rightarrow \infty$, and $\log p /n \rightarrow 0$, where the convergence is in distribution and $r_d \sim N_k(0_{k \times 1}, C_d)$ for some symmetric positive definite matrix $C_d$. 
Let $\Mcal_{d}^*$ be the set of $j$s such that $\lambda_{dj}^*$ is nonzero, then $\Mcal_{d}^* = \{j : (d, j) \in \Mcal^* \}$ and $\Mcal_{d}^{*^c} = \{j : (d, j) \notin \Mcal^*
\}$.  If $A_{{\Bcal \Bcal}}$ denotes a sub-matrix that contains the rows and the columns of matrix $A$ with indices in $\Bcal$, then the block partitioned form of the covariance matrix of $r_d$ in \eqref{eq:td2-4} based on  $\Mcal_{d}^* $ is
\begin{align}
\label{eq:td2-7}
   C_d^* = 
  \begin{bmatrix}
    C_{d_{\Mcal_{d}^* \Mcal_{d}^*}} & C_{d_{\Mcal_{d}^* \Mcal_{d}^{*^c}}}\\
    C_{d_{\Mcal_{d}^{*^c} \Mcal_{d}^*}} & C_{d_{\Mcal_{d}^{*^c} \Mcal_{d}^{*^c}}}
  \end{bmatrix}, \quad (r_{d_{\Mcal^*_d}}, r_{d_{\Mcal^{*^c}_d}})^T \sim N_k \left(0_{k \times 1}, C_d^* \right), \quad (d = 1, \ldots, p),
\end{align}
where $r_{d_{\Mcal^*_d}}$ and $r_{d_{\Mcal^{*^c}_d}}$ include elements of $r_d$ with indices in $\Mcal^*_d$ and  $\Mcal^{*^c}_d$, respectively. Finally, the limiting form of $T_{3dj}$ is found using arguments in \citet{ZouLi08}. If $\lambda_{dj}^* \neq 0$, then 
$\eta_j + p^{-1/2}{|\lambda_{dj}^0|} = \eta_j + p^{-1/2}{|\lambda_{dj}^{*}|} + O_P\{(np)^{-1/2} \}$, ${(np)}^{1/2} \big( 
\big|{p^{-1/2}} {\lambda^*_{dj}} + (np)^{-1/2}{u_{dj}} \big| - \big| p^{-1/2} {\lambda^*_{dj}} \big| \big) = \sgn{(\lambda_{dj}^*)} u_{dj}$, and 
\begin{align*}
  T_{3dj} = \frac{ \big\{n^{-1/2}(\alpha_j + 1)\big\} \, \big[ (np)^{1/2} \big\{ \big| {p^{-1/2}}  {\lambda^*_{dj}}  + {{(np)^{-1/2}}} {u_{dj}} \big| - \big|{p^{-1/2}} {\lambda^*_{dj}} \big| \big\} \big]}{\big\{p^{1/2} \eta_j + |\lambda_{dj}^{*} | + O_P( {{n^{-1/2}}}) \big\}}
 \rightarrow 0
\end{align*}
in probability by Slutsky's theorem and the continuous mapping theorem as $n \rightarrow \infty$, $n \leq p \rightarrow \infty$, and $\log p /n \rightarrow 0$. Similarly, if $\lambda_{dj}^* = 0$, then 
$\eta_j + {p^{-1/2}|\lambda_{dj}^0|} = \eta_j + O_P \{(np)^{-1/2}\}$, 
${(np)^{1/2}} \big( \big| {p^{-1/2}} {\lambda^*_{dj}} + (np)^{-1/2} {u_{dj}} \big| - \big| {p^{-1/2}} \lambda^*_{dj}  \big| \big) = |u_{dj}|$, and
\begin{align}
  T_{3dj} =  \frac{(\alpha_j + 1) \big[(np)^{1/2} \big\{\big|{p^{-1/2}}{\lambda^*_{dj}}   + {{(np)^{-1/2}}} {u_{dj}} \big| - \big| {p^{-1/2}} {\lambda^*_{dj}} \big| \big\} \big]}{\big\{ {(np)^{1/2}} \eta_j + O_P(1) \big\}} \rightarrow
  \begin{cases}
    0, & u_{dj} = 0, \\
    \infty,  & u_{dj} \neq 0,
  \end{cases} \label{dj-eq-0}
\end{align}
in probability by Slutsky's theorem and the continuous mapping theorem as $n \rightarrow \infty$, $n \leq p \rightarrow \infty$, and $\log p /n \rightarrow 0$.

Let $\widehat U_n = \underset{U}{\argmin} \{V_n(U) - V_n(0)\}$, then $\hat \lambda_{dj} / p^{1/2} = \lambda_{dj}^*/ p^{1/2} + {\widehat u_{dj_n}}/{{(np)^{1/2}}}$ or $n^{1/2} (\hat \lambda_{dj} -  \lambda_{dj}^*) = {\widehat u_{dj_n}}$. The limiting forms of 
$T_{1d}$, $T_{2d}$, and $T_{3dj}$ ($d=1, \ldots, p$; $j=1, \ldots, k$), and Slutsky's theorem imply that $V_n(U) - V_n(0)  \goesto V^*(U)$ in distribution for every $U$ as $n \rightarrow \infty$, $n \leq p \rightarrow \infty$, and $\log p /n \rightarrow 0$, where
\begin{align}
   V^*(U) = 
    \begin{cases}
      \sum_{(d, j) \in \Mcal^*} \frac{u_{dj}^2}{2\sigma_{d}^{2^*}} - \sum_{(d, j) \in \Mcal^*} \frac{u_{dj}r_{dj}} {\sigma_{d}^{2^{*}}}, &  u_{dj} = 0\; \text{ for all } (d, j) \notin \Mcal^* , \\
      \infty,  & \text{otherwise.}
    \end{cases}
\end{align}
Since $V_n(U) - V_n(0)$ is convex, the unique minimizer of $V^*(U)$ is
\begin{align}
  U^* \text{ such that  }  u^*_{dj} =  \begin{cases}
    0, & \,  (d, j) \notin \Mcal^*, \\
    r_{dj}, & \,  (d, j) \in \Mcal^*.
    \end{cases} \label{min-u}
\end{align}
Following the epi-convergence results of \citet{Gey94} and \citet{KniFu00}, $\widehat u_{dj_n} \goesto u^*_{dj}$ in distribution ($d=1, \ldots, p$; $j=1, \ldots, k$) as $n \rightarrow \infty$, $n \leq p \rightarrow \infty$, and $\log p /n \rightarrow 0$. Let $\hat \lambda = (\hat \lambda_1^T, \ldots, \hat \lambda_p^T)^T$, $\lambda^* = (\lambda_1^{*^T}, \ldots, \lambda_p^{*^T})^T$, and $\left|A\right|$ be the cardinality of set $A$, then 
\begin{align*}
  n^{1/2} (\hat \lambda_{{\Mcal^*}} - \lambda^*_{\Mcal^*}) \goesto \left( r_{1_{\Mcal^*_1}}, \ldots, r_{p_{\Mcal^*_p}} \right)^T \equiv r_{\Mcal^*}, \quad  \hat \lambda_{{\Mcal^{*^c}}}  \goesto 0_{ \left|\Mcal^{*^c} \right| \times 1} 
\end{align*}
in distribution as $n \goesto \infty$, $n \leq p \rightarrow \infty$, and $\log p /n \rightarrow 0$ 
using \eqref{eq:td2-7}, \eqref{dj-eq-0}, \eqref{min-u}, and $n^{1/2}(\hat \lambda_{dj} -  \lambda_{dj}^*) = \widehat u_{dj_n}\goesto u^*_{dj}$ in distribution ($d=1, \ldots, p$; $j=1, \ldots, k$). Further,
\begin{align*}
  r_{\Mcal^*} \sim N_{\left| \Mcal^* \right|} \left( 0_{\left| \Mcal^* \right| \times 1}, C_{\Mcal^* \Mcal^*} \right), \quad C_{\Mcal^* \Mcal^*}= \mathrm{bdiag}(C_{1_{\Mcal_{1}^* \Mcal_{1}^*} }, \ldots, C_{p_{\Mcal_{p}^* \Mcal_{p}^*}}),
\end{align*}
where $\mathrm{bdiag}(C_{1_{\Mcal_{1}^* \Mcal_{1}^*} }, \ldots, C_{p_{\Mcal_{p}^* \Mcal_{p}^*}})$ is a block diagonal matrix with $C_{1_{\Mcal_{1}^* \Mcal_{1}^*} }, \ldots, C_{p_{\Mcal_{p}^* \Mcal_{p}^*}}$ forming the diagonal blocks.  This proves the asymptotic normality of nonzero loadings.

{Step 2.} We now prove the consistency of $\hat \lambda_{dj}$ ($d=1, \ldots, p$; $j=1, \ldots, k$). For every $(d, j) \in \Mcal^*$, asymptotic normality of $\hat \lambda_{dj}$ implies that $\lambda_{dj} \rightarrow \lambda_{dj}^*$ in probability, so $\text{pr}\{(d, j) \in \hat \Mcal \} \goesto 1$, where $\hat \Mcal$ is the estimated set of the locations of nonzero loadings based on  $\hat \Lambda$. The proof is completed by showing that for all $(\tilde d, \tilde j) \notin \Mcal^*$, $\text{pr}\{(\tilde d, \tilde j) \in \hat \Mcal \} {\goesto} 0$. Let $(\tilde d, \tilde j) \in \hat \Mcal$, then Karush-Kuhn-Tucker optimality condition implies that
\begin{align}
  n^{-1/2} z_{\tilde j}^{0^T} \big( y_{\tilde d} -  Z^0 \hat \lambda_{\tilde d}  \big) = \text{sign}(\hat \lambda_{\tilde d \tilde j}) \frac{\sigma_{\tilde d }^{2^0}(\alpha_{\tilde j} + 1)}{\eta_{\tilde j} {(np)^{1/2}} + O_P(1) }  . \label{kkt}
\end{align}
The right hand side of \eqref{kkt} is unbounded in probability as $n \rightarrow \infty$, $n \leq p \rightarrow \infty$, and $\log p /n \rightarrow 0$ because $(\tilde d, \tilde j) \notin \Mcal^*$. The left hand side of  \eqref{kkt} is
\begin{align}
 n^{1/2} \left(\frac{z_{\tilde j}^{0^T} y_{\tilde d}} {n} - \frac{z_{\tilde j}^{0^T}Z^0 } {n} \lambda_{\tilde d}^{*} \right)  + \frac{z_{\tilde j}^{0^T} Z^0}{n} \left\{ n^{1/2} \left( \lambda_{\tilde d}^{*} - \hat \lambda_{\tilde d}  \right) \right\} .  \label{consist-beta}
\end{align}
Following arguments similar to those used to derive \eqref{eq:td2-4}, the first term in (\ref{consist-beta}) is asymptotically normal. The second term in (\ref{consist-beta}) is also asymptotically normal from asymptotic normality of the estimates of nonzero loadings shown previously. By Slutsky's theorem, the left hand side of \eqref{kkt} is asymptotically normal; therefore, 
 \begin{align}
   \text{pr} \{(\tilde d, \tilde j) \in \hat \Mcal \} \leq \text{pr} \left\{ n^{-1/2} z_{\tilde j}^{0^T} \big( y_{\tilde d} -  Z^0 \hat \lambda_{\tilde d}  \big) = \text{sign}(\hat \lambda_{\tilde d \tilde j}) \frac{\sigma_{\tilde d }^{2^0}(\alpha_{\tilde j} + 1)}{\eta_{\tilde j} {(np)^{1/2}} + O_P(1) } \right\} \goesto 0 
 \end{align}
in probability because asymptotic normality of $n^{-1/2} z_{\tilde j}^{0^T} \big( y_{\tilde d} -  Z^0 \hat \lambda_{\tilde d}  \big) $ implies that it is bounded in probability. This proves the consistency of $\hat \lambda_{dj}$ ($d=1, \ldots, p$; $j=1, \ldots, k$).

\subsection{Proof of asymptotic normality and consistency of estimated $\Sigma$}
 
We now prove asymptotic normality and consistency of $\hat \sigma_{d}^{2}$ ($d=1, \ldots, p$). We first show that $\hat \sigma_{d}^{2}$ is consistent. For the root-$n$ consistent sequence of estimators $\lambda^0_{dj} / p^{1/2}$, ($d=1, \ldots, p$; $j=1, \ldots, k$), Assumption A.5 in the main paper and the continuous mapping theorem imply that if
$n \rightarrow \infty$, $n \leq p \rightarrow \infty$, and $\log p /n \rightarrow 0$, then $L^0 = \{\Omega^{*} + o_{P}(1)\} \{\Omega^{*^{-1}}\Lambda^* + o_{P}(1)\} $, where convergence is element-wise, and 
\begin{align}
   \hat \sigma_{d}^{2} = &\{1 + o(1)\} \left\{ \lambda_d^{*^T} \lambda_d^{*} + o_{P}(1) - 2 \lambda_d^{*^T} \lambda_d^{*} + o_{P}(1) + (\Omega^*)_{dd} + o_{P}(1) \right\}  \nonumber\\
    = &-\lambda_d^{*^T} \lambda_d^{*} + (\Omega^*)_{dd} + o_{P}(1) = \sigma_{d}^{2^*} +  o_{P}(1), \label{con-sig2}
\end{align}
which proves the consistency of $\hat \sigma_{d}^{2}$.

The asymptotic normality of $\hat \sigma_{d}^{2}$ follows from Equation (5.19) and Exercise 5.20 in  \citet{Van00} because the objective for estimating $\hat \sigma_{d}^{2}$ has two continuous derivatives with respect to $\sigma_{d}^2$ for any $Y$ and $\Lambda$. 

\subsection{Lemma required to prove Theorem 3}
\label{necc-lemma}

We use the eigen decomposition of $Y^TY/n$ to impute $\Sigma$ and $Z$ in Equation (3) of the main paper. 
Using the notation of Algorithm 1 in the main paper, impute $\Sigma$ by $\Sigma^0$  and $Z$ by $Z^0$  and let $y=\text{vec}(Y)$, $\lambda = \text{vec}(\Lambda^T)$, $\epsilon = \text{vec}(E^T)$, and $X = I_p \otimes Z^{0} \in \Re^{pn \times pk}$. Then, the hierarchical model for the joint distribution of $y$ and $\lambda$ after scaling Equation (3) in the main paper by $p^{1/2}$ is   
\begin{align}  
  &p^{-1/2} y \mid \lambda \sim N_{np}(X \, p^{-1/2}  \lambda, p^{-1} \Sigma^{0} \otimes I_n), \nonumber \\
  &\lambda \mid \delta, \rho \sim \text{multiscale generalized double Pareto} \{ \alpha_1(\delta), \ldots, \alpha_k(\delta), p^{1/2}\eta_1(\rho), \ldots, p^{1/2}\eta_k(\rho) \}. \label{vec-fa}
\end{align}
The density of the prior for loadings that are estimated to be nonzero in $\Mcal$ is $\prod_{(d, j) \in \Mcal} p_{\text{gdP}}(\lambda_{dj})$, where $p_{\text{gdP}}(\cdot)$ is the density of the generalized double Pareto prior in Section 2.2 of the main paper. The log likelihood of $\lambda_{\Mcal}$ given $\Mcal$ is 
\begin{align}
  \log f_G(y \mid \lambda_{\Mcal}) = &\frac{np \log p }{2} - \frac{np}{2} \frac{\sum_{d=1}^p \log (\Sigma^0)_{dd}}{p} - \frac{np}{2} \log 2 \pi - \frac{1}{2} \sum_{i=1}^n \sum_{d=1}^{p} y_{id}^2/ (\Sigma^0)_{dd} -\nonumber \\
                                   &\frac{n}{2} \sum_{(d, j) \in \Mcal} \lambda^2_{dj} / (\Sigma^0)_{dd}  + n \sum_{(d, j) \in \Mcal} \lambda_{dj} \lambda^0_{dj} / (\Sigma^0)_{dd} \label{llk}
\end{align}
and the log joint density of $y$ and $\lambda_{\Mcal}$ given $\Mcal$ is
\begin{align}
  \log f(y, \lambda_{\Mcal} \mid \delta, \rho) &= \log f_G(y \mid \lambda_{\Mcal}) + \sum_{(d, j) \in \Mcal} \log p_{\text{gdP}}(\lambda_{dj}). \label{xpost}
\end{align}
The following lemma describes the order of $\log f_G(y \mid \lambda_{\Mcal})$ and $\log f(y, \lambda_{\Mcal} \mid \delta, \rho)$ when $\lambda_{\Mcal}$ is replaced by a consistent estimator of $\lambda^*_{\Mcal}$  and $n \rightarrow \infty$, $n \leq p \rightarrow \infty$, and $\log p / n \rightarrow 0$. 
\begin{lemma}
  If $\tilde \lambda_{\Mcal}$ and $\hat \lambda_{\Mcal}$ are root-$n$ consistent estimators of $\lambda^*_{\Mcal}$ and Assumptions A.0--A.7 in the main paper hold, then 
  \begin{align*}
    2\log f_G(y \mid \tilde \lambda_{\Mcal}) / (np \log p) = 2\log f(Y, \hat \lambda_{\Mcal} \mid \delta, \rho) / (np \log p)  = 1 + o_P(1).
  \end{align*}
\end{lemma}
\begin{proof}
  We first show that
  \begin{align*}
    2\log f_G(y \mid \tilde \lambda_{\Mcal}) / (np \log p)    = 1 + o_P(1).
  \end{align*}  
  Using \eqref{llk},  $2 \log f_G(y \mid \tilde \lambda_{\Mcal}) / (np \log p)$
  \begin{align*}
    =& 1 - \frac{1}{\log p} \frac{\sum_{d=1}^p \log (\Sigma^0)_{dd}}{p} - \frac{\log 2 \pi}{\log p}  - \frac{1}{\log p} \frac{ \sum_{i=1}^n \sum_{d=1}^{p}  y_{id}^2/ (\Sigma^0)_{dd} }{np} - \\
    &\frac{\sum_{(d, j) \in \Mcal} \tilde \lambda^2_{dj} / (\Sigma^0)_{dd}}{p \log p}  + 2 \frac{\sum_{(d, j) \in \Mcal} \tilde \lambda_{dj} \lambda^0_{dj} / (\Sigma^0)_{dd}}{p \log p} \\
    =& 1 - \frac{1}{\log p}\frac{\sum_{d=1}^p \log (\Sigma^*)_{dd}}{p} + o_P(1) - o(1) - \frac{1}{\log p} \frac{ \sum_{i=1}^n \sum_{d=1}^{p}  y_{id}^2/ (\Sigma^*)_{dd} }{np} O_P(1) - \\
    &\frac{\sum_{(d, j) \in \Mcal} \lambda^{*^2}_{dj} / (\Sigma^*)_{dd}} {p \log p}  O_P(1) + 2 \frac{\sum_{(d, j) \in \Mcal} \lambda^{2^*}_{dj} / (\Sigma^*)_{dd}}  {p \log p}O_P(1),
  \end{align*}
  where the last equality follows because $(\Sigma^0)_{dd}$ and $\tilde \lambda_{dj}$ are consistent estimators of $(\Sigma^*)_{dd}$ and $\lambda_{dj}^*$ ($d=1, \ldots, p$; $j=1, \ldots, k$). Since $E(y_{id}^2) \leq D_0$ and $D_1 \leq (\Sigma^*)_{dd} \leq D_2$ ($i=1, \ldots, n$; $d=1, \ldots, p$) using Assumption A.1 in the main paper, 
  \begin{align*}
    \sum_{i=1}^n \sum_{d=1}^{p}  (\Sigma^*)^{-1}_{dd} y_{id}^2 /(np) = O_P(1)
  \end{align*}
  by an application of Markov's inequality and
  \begin{align*}
    \log D_1 &\leq \sum_{d=1}^p \log (\Sigma^*)_{dd} / p  \leq \log D_2, \\
    0 \leq \sum_{(d, j) \in \Mcal} (\Sigma^*)^{-1}_{dd} \lambda^{2^*}_{dj} / (p \log p) &\leq \tr({\Omega^{*}}) / (D_1 p \log p) \leq D_0 / (D_1 \log p).
  \end{align*}
  Therefore,
  \begin{align*}
    \frac{2 \log f_G(y \mid \tilde \lambda_{\Mcal}, \Sigma^0)}{np \log p} = 1  + o_P(1) + \frac{O_P(1)}{\log p}  + \frac{O_P(1) }{\log p} = 1 + o_P(1).  
  \end{align*}
  Proceeding similarly, 
  \begin{align*}
    \frac{2 \log f_G(y \mid \hat \lambda_{\Mcal})} {np \log p} = 1 + o_P(1)
  \end{align*}
  using the consistency of $\hat \lambda_{\Mcal}$.

  We complete the proof by showing that 
  \begin{align*}
    \frac{\sum_{(d, j) \in \Mcal} \log p_{\text{gdP}} (\hat \lambda_{dj})}{np \log p} = o_P(1). 
  \end{align*}
  Using the analytic form of $p_{\text{gdP}}$ in \eqref{xpost}, 
  \begin{align}
    \sum_{(d, j) \in \Mcal} \log p_{\text{gdP}}(\hat \lambda_{dj}) = 
    \sum_{(d, j) \in \Mcal} \log \frac{\alpha_j}{p^{1/2} \eta_j} -  \sum_{(d, j) \in \Mcal} (\alpha_j + 1) \log \left(1 + \frac{|\hat \lambda_{dj}|}{p^{1/2} \eta_j } \right). \label{pr1}
  \end{align}
  The first term on the right hand side of \eqref{pr1} after scaling by $np\log p$ is
  \begin{align*}
    \frac{1} {n p \log p} \sum_{(d, j) \in \Mcal} \log \frac{\alpha_j}{p^{1/2} \eta_j}
    &= \frac{1}{p \log p} \sum_{(d, j) \in \Mcal} \left[ \frac{\log \alpha_j} {n}  - \frac{\log \left\{ (np)^{1/2}  \eta_j \right\}}{n} + \frac{\log n}{2n}\right]
    \\
    &= o(1) \frac{|\Mcal|}{p \log p} = o(1) O(1) = o(1).
  \end{align*}
  The last equality follows from Assumption A.5 in the main paper and using conditions that $|\Mcal| \leq p k$ and $k = O(\log p)$. The second term on the right hand side of \eqref{pr1} after scaling by $np\log p$ is
  \begin{align*}
    \frac{1} {n p \log p} \sum_{(d, j) \in \Mcal} (\alpha_j + 1) \log \left(1 + \frac{|\hat \lambda_{dj}|}{p^{1/2} \eta_j } \right)
    =& \frac{1}{p \log p} \sum_{(d, j) \in \Mcal} \frac{\alpha_j + 1}{{n^{1/2}}} \frac{\log \{ {(np)^{1/2}} \eta_j + {n^{1/2}} \hat \lambda_{dj} \}}{{n^{1/2}}} -  \\
    & \frac{1}{p \log p} \sum_{(d, j) \in \Mcal} \frac{\alpha_j + 1}{{n^{1/2}}} \frac{ \{\log {(np)^{1/2}} \eta_j\}}{ {n^{1/2}}}\\
    =& o_P(1) \frac{|\Mcal|}{p \log p} - o(1) \frac{|\Mcal|}{p \log p} = o_P(1).
  \end{align*}
  The last equality follows from Assumption A.5 in the main paper, from consistency of $\hat \lambda_{dj}$, and using conditions that $|\Mcal| \leq p k$ and $k = O(\log p)$. The proof is completed by using \eqref{xpost} to obtain that
  \begin{align*}
    \frac{2\log f(y, \hat \lambda_{\Mcal} \mid \delta, \rho)} {np \log p} = \frac{2\log f_G(y \mid \hat \lambda_{\Mcal})} {np \log p} + \frac{2\sum_{(d, j) \in \Mcal} \log p_{\text{gdP}}(\hat \lambda_{dj})}{np \log p} = 1+ o_P(1).
  \end{align*}
\end{proof}

\subsection{Proof of Theorem 3}
\label{bma-y}

The proof consists of three steps: derive the asymptotic form of $\log \pi_{\Mcal}$; show that $-2 \log \pi_{\Mcal} / \textsc{ebic}_{\gamma}(\Mcal) = 1 + o_P(1)$; and show that the sufficient condition for model selection consistency of \textsc{ebic}$_{\gamma}$($\Mcal$) holds under the assumptions of Theorem 3 in the main paper. 

We use the following notation for ease of presentation. If $\Bcal$ is a set of indices and $X$ is a matrix, then $X_{\Bcal}$ is a sub-matrix that contains columns of $X$ with indices in $\Bcal$ and $X_{\Bcal, \Bcal}$ is a sub-matrix that contains rows and columns of $X$ with indices in $\Bcal$. 

Step 1.
Using \eqref{vec-fa}, the density of the prior for loadings that are estimated to be nonzero in $\Mcal$ is $\prod_{(d, j) \in \Mcal} p_{\text{gdP}}(\lambda_{dj})$; see Section \ref{necc-lemma} also. Use the Gaussian scale mixture  representation for the density of generalized double Pareto prior to write $|\lambda_{dj}|$ in form of differentiable functions when $\lambda_{dj} \neq 0$; see the equation for E-step in Section 4.4.1 of \citet{ArmDunLee11} for details related to the Gaussian scale mixture representation for the generalized double Pareto density. Define the  diagonal matrix $D$ as
\begin{align*}
  D = \frac{d^2 \log \big\{\prod_{(d, j) \in \Mcal} p_{\text{gdP}}(\lambda_{dj})\big\} } {d \lambda_{\Mcal} d \lambda_{\Mcal}^T}, 
\end{align*}
and let 
\begin{align*}
  D_{(d, j), (d, j)} = \frac{\alpha_j(\delta) + 1}{\{p^{1/2} \eta_j(\rho) + |\lambda_{dj}|\}^2}
\end{align*}
be the diagonal element of $D$ corresponding to $\lambda_{dj}$ such that $(d, j) \in \Mcal$. If $f(y, \lambda_{\Mcal} \mid \delta, \rho)$ is the joint density of $y$ and $\lambda_{\Mcal}$ defined using \eqref{vec-fa}, then define another diagonal matrix $H_{\Mcal}$ as
\begin{align}
  H_{\Mcal} = -\frac{d^2 \log f(y, \lambda_{\Mcal} \mid \delta, \rho)}{d \lambda_{\Mcal} d \lambda_{\Mcal}^T} = n (\Sigma^{0^{-1}} \otimes I_n)_{\Mcal, \Mcal} - D. \label{hess-1}
\end{align}
If $\hat H_{\Mcal}$ represents $H_{\Mcal}$ in \eqref{hess-1} evaluated at $\hat \lambda_{\Mcal}$, then the diagonal element of $\hat H_{\Mcal}$ that corresponds to the index $(d, j) \in \Mcal$ is
\begin{align}
  \label{comp-inf}
  \frac{n}{\sigma^{2^{0}}_{d}}  - \frac{\alpha_j(\delta)  + 1 } {\{p^{1/2}  \eta_j(\rho) + |\hat \lambda_{dj}|\}^2}     =  \begin{cases}
    \frac{n}{\sigma^{2^{*}}_{d}} \left\{ 1 + o_P(n^{-1/2}) \right\}, & (d, j) \in \Mcal^*, \\
    \frac{n}{\sigma^{2^{*}}_{d}} \left\{ 1 + o_P(n^{1/2}) \right\}, &  (d, j) \notin \Mcal^*. 
    \end{cases} 
\end{align}
The equality in \eqref{comp-inf} follows because $\hat \lambda_{dj} = \lambda^*_{dj} + o_P(n^{-1/2})$, $\sigma_{d}^{2^{0}} = \sigma_{d}^{2^{*}} + o_P(n^{-1/2})$, and $\alpha_j(\delta) = o({n^{1/2}})$ from Theorem 2 in the main paper and Assumptions A.0--A.6 in the main paper.  The posterior probability of $\Mcal$, denoted as $\pi_{\Mcal}$, equals
\begin{align}
  \label{bma-1}
   \text{pr}(\Mcal \mid Y, \delta, \rho) \propto m(Y \mid \Mcal) \;\, \text{pr}(\Mcal \mid \delta,\rho) ,
\end{align}
where $m(Y \mid \Mcal)$ is the marginal likelihood of the factor model in \eqref{vec-fa} with the locations of nonzero loadings contained in the set $\Mcal$, $m(y \mid \Mcal) = \int f(Y, \lambda_{\Mcal} \mid \delta, \rho) \, d \lambda_{\Mcal}$, and $\text{pr}(\Mcal \mid \delta,\rho)$ is prior defined in Equation (9) in the main paper. Using Laplace approximation and \eqref{comp-inf},
\begin{align}
  2 \log m(Y \mid \Mcal) &= 2 \log f(Y, \hat \lambda_{\Mcal} \mid \delta, \rho) - |\Mcal| \log n [1 + \{c + o_P(\log n)\} / \log n] , \label{lap1}
\end{align}
where $c =  \log (2 \pi) + \sum_{(d, j) \in \Mcal} \sigma_{d}^{2^*} /  |\Mcal|=O(1)$ using Assumption A.1 in the main paper. Further, using \eqref{bma-1},
\begin{align*}
  -2 \log \pi_{\Mcal} = -2 \log m(Y \mid \Mcal) - 2 \log \text{pr}(\Mcal \mid \delta, \rho).
\end{align*}
Sterling's approximation and Theorem 2 imply that $\log \text{pr}(\Mcal \mid \delta, \rho) = - |\Mcal| \log (pk) \{1 + o_P(1)\}$; therefore, the previous equation after using \eqref{lap1} reduces to
\begin{align}
  -2 \log \pi_{\Mcal}  &= -2 \log f(Y, \hat \lambda_{\Mcal} \mid \delta, \rho) + |\Mcal| \{\log n  + 2 \log (pk)\} \{1 + o_P(1)\}.
  \label{wt1}
\end{align}

{Step 2.} The definition of $\textsc{ebic}_{\gamma}(\Mcal)$ in \citet{CheChe08} for regression models implies that
\begin{align}
  \textsc{ebic}_{\gamma}(\Mcal) &= - 2 \log f_G(y \mid \tilde \lambda_{\Mcal}) + |\Mcal| \left\{ \log (np) + 2 \gamma \log (pk) \right\}, \nonumber \\
  &= - 2 \log f_G(y \mid \tilde \lambda_{\Mcal}) + |\Mcal| \left\{ \log n + (2 \gamma + 1) \log p \right\}\{1 + o_P(1)\}, \label{ebic-1}
\end{align}
where $\tilde \lambda_{dj}$ is a root-$n$ consistent estimate of $\lambda^*_{dj}$  ($d=1, \ldots, p$; $j=1, \ldots, k$) in \eqref{vec-fa}, $f_G(y \mid \lambda_{\Mcal})$ is the Gaussian likelihood defined using \eqref{vec-fa}, and $0 < \gamma < 1$ is a tuning parameter such that $\gamma > 1 - 1 /(2 \kappa)$. Lemma \ref{necc-lemma} implies that there exists a universal constant $b^*$ such that
\begin{align}
  -2\log f_G(y \mid \tilde \lambda_{\Mcal}) / (np\log p) = -2\log f(Y, \hat \lambda_{\Mcal} \mid \delta, \rho) / (np \log p)  = b^* + o_P(1). \label{ord-llk}
\end{align}
Let $r = -2 \log \pi_{\Mcal}/\textsc{ebic}_{\gamma}(\Mcal) $. Then, Theorem 2 in the main paper and \eqref{ord-llk} imply that
\begin{align}
r &=\frac{ -2 \log f(Y, \hat \lambda_{\Mcal} \mid \delta, \rho) /(np\log p)   + |\Mcal| \{\log n  + 2 \log (pk)\} / (np\log p) \{1 + o_P(1)\}}{- 2 \log f_G(y \mid \tilde \lambda_{\Mcal}) / (np\log p) + |\Mcal| \left\{ \log n + (2 \gamma + 1) \log p \right\} / (np\log p) \{1 + o_P(1)\}}  \nonumber\\
  &=\frac{b^* + o_P(1) +\{|\Mcal^*| + o_P(1)\} o_P(1)}{b^* + o_P(1) +\{|\Mcal^*| + o_P(1)\} o_P(1)} = 1 + o_P(1). \label{ebic-wt}
\end{align}

{Step 3.} Let $l$ be an upper bound on $k^*$ in \eqref{vec-fa} such that $X \in \Re^{pn \times pl}$. If $(np)^{-1} X^T X$ has positive eigen values for any $l$ such that $k \leq l \leq 2k$, $\Mcal \neq \Mcal^* $, and $|\Mcal| \in \{1, \ldots, pk\}$, then uniformly for any such $\Mcal$ there is a universal positive constant $C_0$ and a positive constant $C_{\Mcal}$ depending on $\Mcal$ such that
\begin{align}
  \label{eq:ebic-suff}
  \textsc{ebic}_{\gamma}(\Mcal) - \textsc{ebic}_{\gamma}(\Mcal^*) \geq  \begin{cases}
    C_0 \log n \{1 + o_P(1)\}, & \Mcal^* \subset \Mcal, \\
    C_{\Mcal} \log n, & \text{otherwise}; \\
    \end{cases} 
\end{align}
see the definition of asymptotic identifiability condition on pages 762--763 in \citet{CheChe08} and the proof of Theorem 1 in \citet{CheChe08}. Using \eqref{ebic-wt} and \eqref{eq:ebic-suff}, $  2 \log (\pi_{\Mcal^*} / \pi_{\Mcal})  = \{\textsc{ebic}_{\gamma}(\Mcal) - \textsc{ebic}_{\gamma}(\Mcal^*)\} \{1 + o_P(1)\} \rightarrow \infty$ as $n \rightarrow \infty$ for any $\Mcal $ such that $\Mcal \neq \Mcal^*$ and $|\Mcal| \in \{1, \ldots, pk\}$. The proof is completed by showing $(np)^{-1} X^T X$ has positive eigen values for any $l$ such that $k \leq l \leq 2k$. Assumption A.7 implies that $Y^T Y/n$ has at least $2k$ positive eigen values, so $(np)^{-1} X^T X = I_p \otimes (np)^{-1} Z^{0^T} Z^{0} = I_p \otimes  p^{-1} I_l$, which has $pl$ positive eigenvalues equal to $p^{-1} > 0$ for any $k \leq l \leq 2k$.

\section{Microarray data analysis}
\label{sec:appl-expand-fact}

The AGEMAP data \citep{Zahetal07} were obtained from \url{http://statweb.stanford.edu/~owen/data/AGEMAP/}. 

The $\delta$-$\rho$ grid in expandable factor analysis had 20 different $\delta$ and 20 different $\rho$ values: $\delta_i = 10^{a_i}$, where $a_i =\log_{10} 2 + (i - 1)  (\log_{10} 10 - \log_{10} 2) / 20$ ($i=1, \ldots, 20$), and $\rho_i = 10^{b_i}$, where $b_i =\log_{10} 10^{-3} + (i - 1)  (\log 10^6 - \log 10^{-3}) / 20$ ($i=1, \ldots, 20$). Our estimation algorithm estimated $\Lambda$ at grid points ($\delta_r$, $\rho_s$) ($r=1, \ldots, 20$; $s=1, \ldots, 20$). The results of our estimation algorithm were stable in that the estimated rank of $\Lambda$ was the same at most points on the $\delta$-$\rho$ grid across 10 folds of cross-validation (Table \ref{tab:agemap-rnk}).

\begin{table}[h]
  \def~{\hphantom{0}}
  \caption{\it Estimated rank of loadings matrix in AGEMAP data analysis across $\delta$-$\rho$ grid. The results are averaged over 10 folds of cross-validation and the maximum Monte Carlo error is 0$\cdot$52 across the 10 folds.} 
  {\tiny
    \begin{tabular}{rrrrrrrrrrrrrrrrrrrrr}
      & $\delta_1$ & $\delta_2$ & $\delta_3$ & $\delta_4$ & $\delta_5$ & $\delta_6$ & $\delta_7$ & $\delta_8$ & $\delta_9$ & $\delta_{10}$ & $\delta_{11}$ & $\delta_{12}$ & $\delta_{13}$ & $\delta_{14}$ & $\delta_{15}$ & $\delta_{16}$ & $\delta_{17}$ & $\delta_{18}$ & $\delta_{19}$ & $\delta_{20}$ \\ 
      $\rho_{20}$ & 10 & 10 & 10 & 10 & 10 & 10 & 10 &  9 &  9 &  8 &  8 &  7 &  7 &  7 &  6 &  6 &  6 &  6 &  6 &  5 \\ 
      $\rho_{19}$ & 10 & 10 & 10 & 10 & 10 & 10 &  9 &  9 &  8 &  8 &  7 &  7 &  6 &  6 &  6 &  6 &  6 &  5 &  5 &  5 \\ 
      $\rho_{18}$ & 10 & 10 & 10 & 10 &  9 &  9 &  8 &  8 &  7 &  7 &  6 &  6 &  6 &  6 &  6 &  5 &  5 &  5 &  5 &  4 \\ 
      $\rho_{17}$ & 10 & 10 & 10 &  9 &  9 &  8 &  7 &  7 &  7 &  6 &  6 &  6 &  6 &  5 &  5 &  5 &  5 &  4 &  4 &  4 \\ 
      $\rho_{16}$ & 10 & 10 &  9 &  8 &  8 &  7 &  7 &  6 &  6 &  6 &  6 &  5 &  5 &  5 &  4 &  4 &  4 &  4 &  4 &  4 \\ 
      $\rho_{15}$ &  10 &  9 &  8 &  7 &  7 &  6 &  6 &  6 &  6 &  5 &  5 &  4 &  4 &  4 &  4 &  4 &  4 &  4 &  4 &  3 \\ 
      $\rho_{14}$ &  8 &  7 &  7 &  6 &  6 &  6 &  6 &  5 &  4 &  4 &  4 &  4 &  4 &  4 &  4 &  4 &  3 &  3 &  3 &  3 \\ 
      $\rho_{13}$ &  7 &  6 &  6 &  6 &  6 &  5 &  4 &  4 &  4 &  4 &  4 &  4 &  4 &  3 &  3 &  3 &  3 &  3 &  3 &  3 \\ 
      $\rho_{12}$ &  6 &  6 &  5 &  4 &  4 &  4 &  4 &  4 &  4 &  4 &  3 &  3 &  3 &  3 &  3 &  3 &  2 &  2 &  2 &  2 \\ 
      $\rho_{11}$ &  5 &  4 &  4 &  4 &  4 &  4 &  3 &  3 &  3 &  3 &  3 &  3 &  2 &  2 &  2 &  2 &  2 &  2 &  2 &  2 \\ 
      $\rho_{10}$ &  4 &  4 &  4 &  3 &  3 &  3 &  3 &  3 &  2 &  2 &  2 &  2 &  2 &  2 &  2 &  2 &  2 &  2 &  2 &  2 \\ 
      $\rho_{9}$ &  3 &  3 &  2 &  2 &  2 &  2 &  2 &  2 &  2 &  2 &  2 &  2 &  2 &  2 &  1 &  1 &  1 &  1 &  1 &  1 \\ 
      $\rho_{8}$ &  2 &  2 &  2 &  2 &  2 &  2 &  2 &  1 &  1 &  1 &  1 &  1 &  1 &  1 &  1 &  1 &  1 &  1 &  1 &  1 \\ 
      $\rho_{7}$ &  1 &  1 &  1 &  1 &  1 &  1 &  1 &  1 &  1 &  1 &  1 &  1 &  1 &  1 &  1 &  1 &  1 &  1 &  1 &  1 \\ 
      $\rho_{6}$ &  1 &  1 &  1 &  1 &  1 &  1 &  1 &  0 &  0 &  0 &  0 &  0 &  0 &  0 &  0 &  0 &  0 &  0 &  0 &  0 \\ 
      $\rho_{5}$ &  0 &  0 &  0 &  0 &  0 &  0 &  0 &  0 &  0 &  0 &  0 &  0 &  0 &  0 &  0 &  0 &  0 &  0 &  0 &  0 \\ 
      $\rho_{4}$ &  0 &  0 &  0 &  0 &  0 &  0 &  0 &  0 &  0 &  0 &  0 &  0 &  0 &  0 &  0 &  0 &  0 &  0 &  0 &  0 \\ 
      $\rho_{3}$ &  0 &  0 &  0 &  0 &  0 &  0 &  0 &  0 &  0 &  0 &  0 &  0 &  0 &  0 &  0 &  0 &  0 &  0 &  0 &  0 \\ 
      $\rho_{2}$ &  0 &  0 &  0 &  0 &  0 &  0 &  0 &  0 &  0 &  0 &  0 &  0 &  0 &  0 &  0 &  0 &  0 &  0 &  0 &  0 \\ 
      $\rho_{1}$ &  0 &  0 &  0 &  0 &  0 &  0 &  0 &  0 &  0 &  0 &  0 &  0 &  0 &  0 &  0 &  0 &  0 &  0 &  0 &  0 \\ 
    \end{tabular}
  }%
  \label{tab:agemap-rnk}
\end{table}

\begin{figure}[h]
  \subfloat[Sparse principal components]{
    \includegraphics[width=0.68\textwidth]{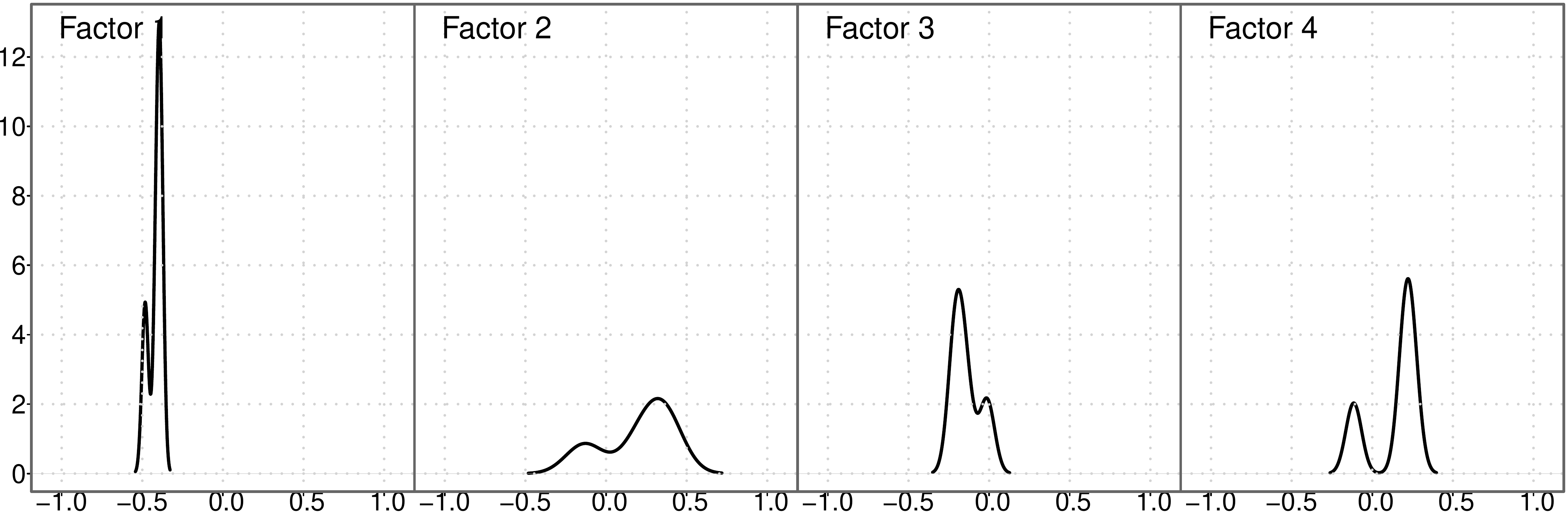}
    \label{fig:cer-spc}}
  \subfloat[Expandable factor analysis]{
    \includegraphics[scale=0.145]{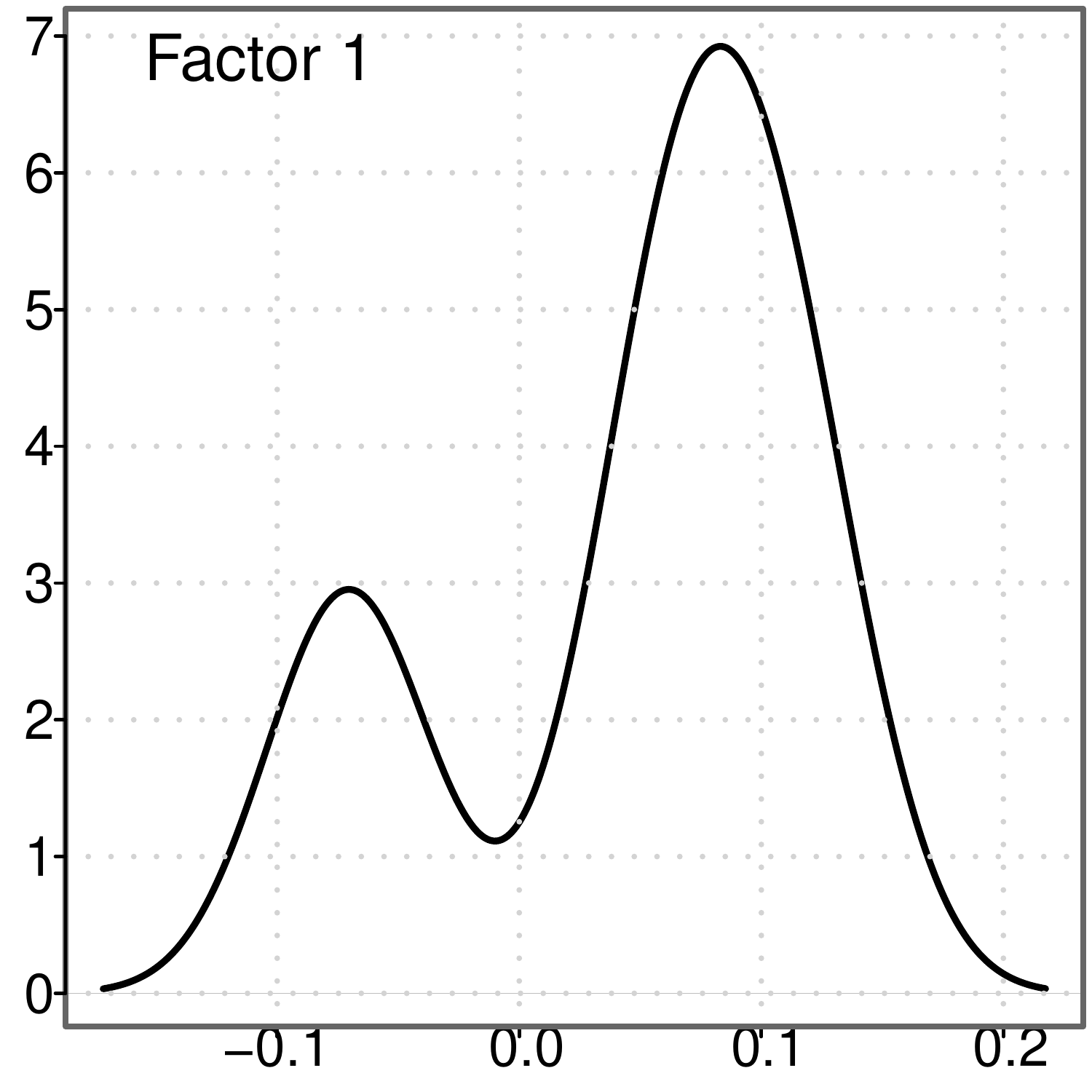}
    \label{fig:cer-xfa}}
\caption{Density plots for the estimated factors in a test data for cerebrum tissue samples.}
\label{fig:fact-cer}
\end{figure}

\begin{figure}[h]
  \subfloat[Sparse principal components]{
    \includegraphics[width=0.68\textwidth]{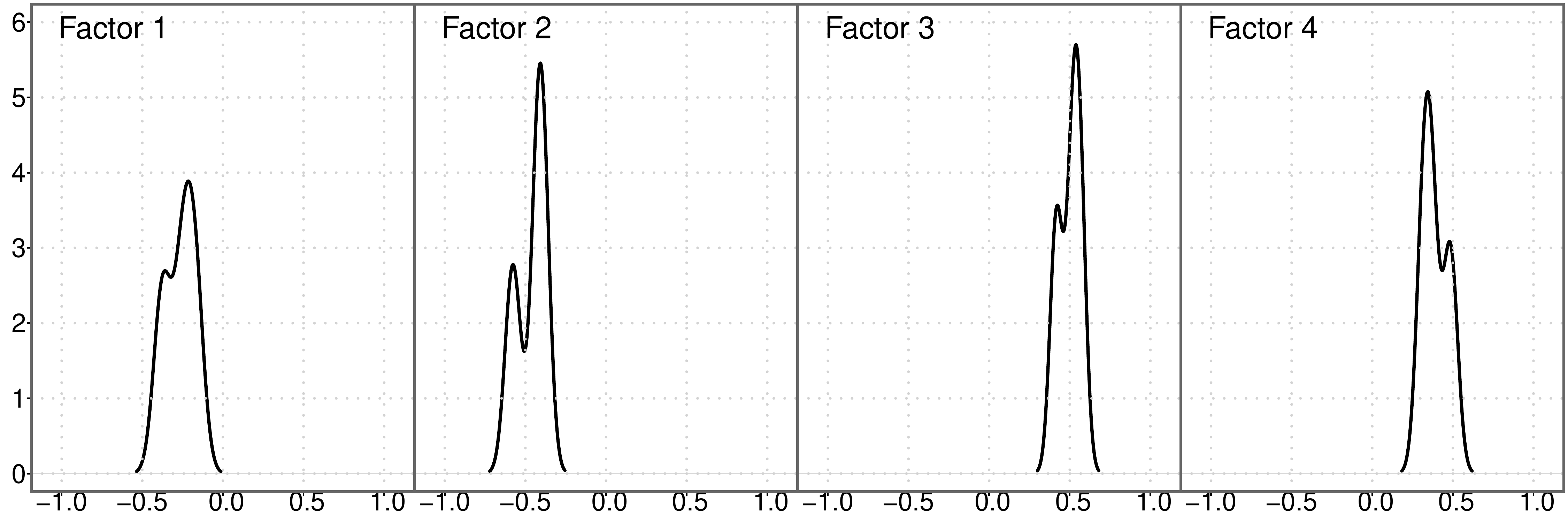}
    \label{fig:bel-spc}}
  \subfloat[Expandable factor analysis]{
    \includegraphics[scale=0.145]{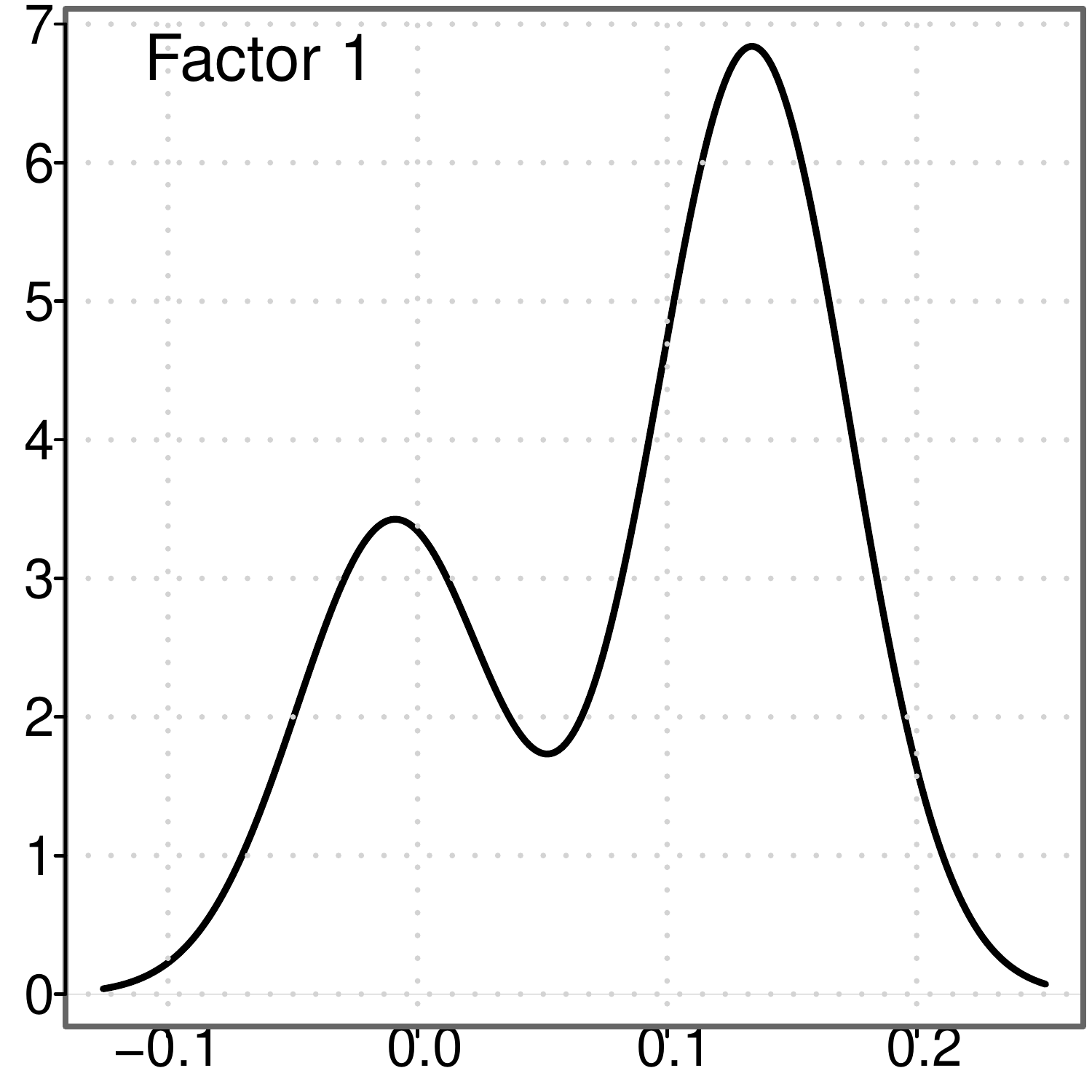}
    \label{fig:bel-xfa}}
\caption{Density plots for the estimated factors in a test data for cerebellum tissue samples.}
\label{fig:fact-bel}
\end{figure}

\end{document}